\documentclass[fleqn,usenatbib]{mnras}

\usepackage{newtxtext,newtxmath}
\usepackage[T1]{fontenc}
\usepackage{ae,aecompl}
\usepackage{adjustbox}
\usepackage{amsfonts}
\usepackage{amsmath}
\usepackage{graphicx}	
\usepackage{tabularx}
\usepackage{verbatim}
\usepackage{xspace}
\usepackage{adjustbox}
\usepackage{physics}
\usepackage{comment}
\usepackage{rotating, lipsum, babel}

\usepackage{natbib}
\usepackage{caption}
\usepackage{hyperref}
\usepackage{url}

\newcommand{\pc}{\,\mathrm{pc}}
\newcommand{\Msun}{\,\mathrm{M}_{\odot}}
\newcommand{\kpc}{\,\mathrm{kpc}}
\newcommand{\Gyr}{\,\mathrm{Gyr}}

\newcommand{\Myr}{\,\mathrm{Myr}}

\newcommand{\kms}{\,\mathrm{km\,s}^{-1}}

\newcommand{\Rc}{R_\mathrm{c}}

\newcommand{\Md}{M_\mathrm{d}}

\newcommand{\phid}{\phi_\mathrm{disc}}
\newcommand{\phih}{\phi_\mathrm{halo}}

\newcommand{\Vinfty}{V_\infty}

\newcommand{\Nbody}{$N$-body\xspace}

\newcommand{\secref}[1]{Section~\ref{#1}}
\newcommand{\figref}[1]{Figure~\ref{#1}}
\newcommand{\tabref}[1]{Table~\ref{#1}}

\title[The impact of IMF on the evolution of DSCs]{The impact of a top-heavy IMF on the formation and evolution of dark star clusters}

\author[Rostami Shirazi et al.]{
Ali Rostami Shirazi,$^{1}$ 
Hosein Haghi, $^{1,4}$\thanks{E-mail: haghi@iasbs.ac.ir}
Akram Hasani Zonoozi, $^{1}$
Ahmad Farhani Asl$^{1}$\\
\newauthor and 
Pavel Kroupa $^{2,3}$
\\
$^{1}$Department of Physics, Institute for Advanced Studies in Basic Sciences (IASBS), 444 Prof. Sobouti Blvd., Zanjan 45137-66731, Iran\\
$^{2}$Helmholtz-Institut f\"ur Strahlen-und Kernphysik (HISKP), Universit\"at Bonn, Nussallee 14-16, D-53115 Bonn, Germany\\
$^{3}$Charles University in Prague, Faculty of Mathematics and Physics, Astronomical Institute, V Hole\v{s}ovi\v{c}k\'ach 2, CZ-180 00 \\
Praha 8, Czech Republic\\
$^{4}$School of Astronomy, Institute for Research in Fundamental Sciences (IPM), PO Box 19395 - 5531, Tehran, Iran\\}

\date{Accepted XXX. Received YYY; in original form ZZZ}

\pubyear{2023}

\begin{document}
\label{firstpage}
\pagerange{\pageref{firstpage}--\pageref{lastpage}}
\maketitle


\begin{abstract}

The Spitzer instability leads to the formation of a black hole sub-system (BHSub) at the center of a star cluster providing energy to luminous stars (LSs) and increasing their rate of evaporation. When the self-depletion time of the BHSub exceeds the evaporation time of the LSs, a dark star cluster (DSC) will appear. Using the NBODY7 code, we performed a comprehensive set of direct \Nbody simulations over a wide range of initial conditions to study the pure effect of the top-heaviness of the IMF on the formation of the DSC phase. In the Galactic tidal field, top-heavy IMFs lead to the fast evaporation of LSs and the formation of DSCs. Therefore, DSCs can be present even in the outer region of the Milky Way (MW). To successfully transition to the DSC phase, the MW Globular Clusters (GCs) must possess an initial BH mass fraction of $\widetilde{\mathit{M}}_\mathrm{BH}(0)>0.05$. For star clusters with $\widetilde{\mathit{M}}_\mathrm{BH}(0)>0.08$, the DSC phase will be created for any given initial density of the cluster and Galactocentric distance. The duration of the cluster’s lifetime spent in the DSC phase shows a negative (positive) correlation with the initial density, and Galactocentric distance of the star cluster if $\widetilde{\mathit{M}}_\mathrm{BH}(0)\leq 0.12$ ($\widetilde{\mathit{M}}_\mathrm{BH}(0)\geq 0.15$). Considering the canonical IMF, it is unlikely for any MW GCs to enter the DSC phase. We discuss the BH retention fraction in view of the observed properties of the GCs of the MW.

\end{abstract}

\begin{keywords}
galaxies: star clusters general - globular clusters: general - stars: luminosity function, mass function - stars: black holes - methods: numerical
\end{keywords}

\section{Introduction}\label{sec:Intro}

Star clusters have been recognized as one of the key environments where black holes (BHs) are formed. However, it was initially believed that BHs experience a natal kick during their supernova explosion, which could accelerate them over the escape velocity, resulting in almost all BHs leaving the cluster as soon as they are formed. The uncertainty about the efficiency of the natal kicks has led to doubts about the number of BHs initially remaining in star clusters. Observations of Galactic low-mass X-ray binaries provide some insight into possible BH natal kicks. Integrating the trajectories of low-mass X-ray binaries containing BHs within the Milky Way (MW) reveals that while the observational parameters of some systems could be explained with no or small natal kicks, others are better described when considering a relatively large natal kick \citep{Repetto2012, Repetto2015, Mandel2016, Repetto2017}. 

Over the last decade, there has been a significant change in our understanding of BHs in old globular clusters (GCs).  \citet{Maccarone2007} identified, for the first time,  a BH X-ray binary candidate inside a GC in the galaxy NGC 4472. Several more BH candidates have also been discovered in extragalactic GCs \citep{Shih2010, Barnard2011, Maccarone2011, Saracino2022}. \citet{Strader2012} also discovered two BHs within the Galactic GC M22 by radio observations, marking the first time a MW GC had ever shown signs of having BH candidates. Considering that these BHs are accreting from white dwarf (WD) companions, and by using formation and survival rates calculated by \citet{Ivanova2010}, about 5–100 BHs are thought to be present in M22. Furthermore, several BH candidates have also been discovered in other MW GCs \citep{Chomiuk2013, Miller-Jones2015, Giesers2018}.

It has been shown in several theoretical studies that GCs might actually be able to retain significant numbers of BHs \citep{Breen2013, Morscher2013, Morscher2015, Pavlik2018, Longwang2020}. By investigating the evolution of two-component clusters consisting of a population of BHs co-existing within a background cluster of low-mass stars, \citet{Breen2013} found that the dynamical ejection rate of BHs is lower than previously thought.  They used analytic calculations and direct \Nbody  simulations, which both suggest that the exchange of energy between the BH-subsystem (BHSub) and the other stars is ultimately controlled by the entire cluster. As a result, the rate of energy production in BHSub, as well as its depletion rate, is also regulated by the whole cluster. In other words, the dynamical evolution timescale of BHSub follows the evolutionary timescale of the whole cluster.  They also concluded that a BHSub can survive roughly for $10\times t_\mathrm{rh}$ (where $t_\mathrm{rh}$ is the half-mass radius relaxation time). Therefore, if $t_\mathrm{rh}$ is long enough, it is expected that MW GCs still host BHs.  \citet{Breen2013} suggested that the collapse of the visible core occurs approximately when all BHs left the cluster. Since only about $20$ per cent of the MW GCs are identified as core collapsed \citep{Djorgovski1986}, it is possible that up to 80 per cent of the Galactic GCs still have populations of BHs.

Several studies, employing direct \Nbody simulations, have shown that retaining BHs within certain clusters is necessary for reproducing their observable evidence. For instance, \citet{Mackey2008} suggested that the observed increase in core radius with age in star clusters of the Large Magellanic Cloud could be explained by a significant retention fraction of BHs because a population of heavy dark remnants inflates the core radius measured from the visible stars in the cluster \citep{Merritt2004}. \citet{Peuten2016} showed that models with a 50 per cent BH remnant fraction can reproduce the lack of mass segregation within  NGC 6101. In a recent study, \citet{Gieles2021} demonstrated that the presence of a BH population within the Palomar 5 cluster, accounting for approximately 20 per cent of the cluster's present-day mass, can give rise to the extended tidal tails and large half-light radius observed in the cluster. Furthermore, \citet{Torniamenti2023} compared density profiles of direct \Nbody models with the \emph{Gaia} data of the Hyades open cluster and concluded that the observations are best reproduced by models with 2-3 BHs at present. The need to have a population of dark heavy remnants has been also put forward to interpret the central cusp in the velocity dispersion and surface brightness profiles of $\omega \ \mathrm{Cen}$. \citet{Zocchi2019}  fitted the velocity dispersion profile of $\omega \ \mathrm{Cen}$ using a modeled cluster that contained segregated BHs. Moreover, \cite{Baumgardt2019} found that a model with 4.6 per cent of the mass of $\omega \ \mathrm{Cen}$ in a centrally concentrated cluster of BHs can fit all available data.

In addition to the researches focused on observational constraints of BH populations in specific clusters, recent works provide broader and more statistically robust constraints on the present-day BH content in numerous MW GCs. Various studies have been conducted using different methods, like the best-fitting multi-mass models, evaluations of visible mass segregation, and assessments of the central surface brightness of several GCs \citep{Askar2018,Weatherford2018,Weatherford2020,Dickson2023}. These studies suggest that a moderate fraction of the present-day total mass in the form of BHs is generally necessary to explain the observations. They identified massive ($>10^5 \Msun$) GCs retaining especially large populations of BHs, of a total masses exceeding $10^3 \Msun$ per cluster.

As another observational example, one can refer to star cluster IRS 13E, which is an extremely compact stellar association of a few young, massive stars that are located close to the Galactic center and have survived the extreme tidal field by being bound by an invisible mass \citep{Maillard2004} despite having relatively high-velocity ($\simeq200 \kms$) stars \citep{Fritz2010}. \citet{banerjee2011} showed that the dark component could be an ensemble of stellar-mass BHs. They predicted that at distances close to the Galactic centre, rapid tidal stripping of star clusters by the strong tidal field can expose their BHSub, which may host a few orbiting stars. This occurs when the evaporation timescale of stars from the outer regions of the cluster is shorter than the encounter-driven self-depletion timescale of its central BHSub. Such clusters appear as highly super-virial star clusters with a large dynamical mass-to-light ($M_\mathrm{dyn}/L$) ratio. These objects belong to a type of compact stellar populations that \citet{banerjee2011} called “dark star clusters” (hereafter DSCs).

Some observational evidence has been reported in the past few years for clusters that could be potential candidates for DSCs due to their high $M_\mathrm{dyn}/L$ ratios. For instance, \citet{Taylor2015} measured the dynamical properties of 125 compact clusters in the giant elliptical Galaxy NGC 5128. However, they could not identify any cluster with unusual kinematic properties in the dynamical mass range $10^5<M_\mathrm{dyn}/\Msun<10^6$. For $M_\mathrm{dyn}/\Msun>10^6$ they observed two distinct sequences of star clusters in the $M_\mathrm{dyn}/L\propto M_\mathrm{dyn}^{\alpha}$ plane. The sequence described by the slope $\alpha=0.33\pm0.04$ exhibited $M_\mathrm{dyn}/L<10\Msun/L_\odot$, while the sequence described by $\alpha=0.79\pm0.04$  exhibited $10\Msun/L_\odot<M_\mathrm{dyn}/L<70\Msun/L_\odot$, which is termed the DSC sequence.

The initial mass function (IMF) of stars within an embedded cluster is also one of the most important initial conditions that play a significant role in star cluster evolution. Most studies of resolved stellar populations in the disk of the MW showed that stars form following an IMF that has a universal form \citep{Bastian2010, kroupa2001, Kroupa2002}, which is referred to as the "canonical" IMF. The shape of the stellar IMF of a star cluster near its upper mass limit is a focal topic of investigation as it determines the high-mass stellar content and hence the dynamics of the cluster in its embedded phase. Several observational and theoretical studies suggest that the IMF slope for massive stars in GCs depends on the initial cloud density and metallicity ($Z$), such that the IMF becomes increasingly top-heavy with decreasing $Z$ and increasing gas density of the forming object \citep{MarksMichael2012}. The need for a top-heavy IMF also has been put forward to explain the observed trend of $Z$ and $M_\mathrm{dyn}/L$ ratio found in the M31 GC system, which shows a discrepancy with the stellar population synthesis (SPS) prediction \citep{Zonoozi2016, Haghi2017}. If this were to be the case, then the evolution of GCs is likely to be significantly different from the case of the canonical IMF due to the different mass-loss rates and number of BHs formed,  consequently affecting our understanding of GC survival.

In this paper, we will explore the formation of DSCs for star clusters starting with a top-heavy IMF. We want to calculate a comprehensive grid of models over a wide range of initial half-mass radii ($r_\mathrm{h,i}$), Galactocentric distances ($R_\mathrm{G}$), $Z$, and varying the IMF-slope in the high-mass range to investigate the starting time of the DSC phase. We aim to shed light on the pure effect of the top-heaviness of the IMF as well as different values of the compact remnant retention fraction on the starting time and the lifetime of the DSC phase by means of direct \Nbody simulations. This paper is organized as follows: In \secref{sec:method}, we describe the initial setup of the \Nbody models and our simulation method. In \secref{sec:WH-DSC}, we describe the dynamical process of BHSub and features of the DSC phase. The main results are presented in \secref{sec:result}. Finally, \secref{sec:conclusion} consists of a discussion and the conclusions.

\section{Description of the models}\label{sec:method}

\begin{table}
	\centering
	\begin{tabular}{cccccccc}
		\hline
		  & &  & & Set A &  & &\\ 
		\hline  
		Model & $ r_\mathrm{h,i}\ $ & $ Z\ $ & $ \alpha_3 $ & $ R_\mathrm{G}\ $ & $\tau_\mathrm{DSC}$ & $\tau_\mathrm{cluster}$ & ${\tau_\mathrm{DSC}/}$ \\ 
		& $(\pc)$ & $(Z_\odot)$ & & $(\kpc)$ & $(\Myr)$& $(\Myr)$ &  ${\tau_\mathrm{cluster}}$
		\\
		\hline 

        A1  & 1 & 0.25 & 2.3 & 2 & 61 & 1623 &0.037 \\
        A2  & 1 & 0.25 & 2.3 & 3 & 0 &4509 & 0 \\
        A3  & 3 & 0.25 & 2.3 & 2 & 130 &920 &0.141 \\
        A4  & 3 & 0.25 & 2.3 & 3 & 307 &2376 &0.129 \\
        A5  & 3 & 0.25 & 2.3 & 4 & 340 &4490 &0.075 \\
        A6  & 3 & 0.25 & 2.3 & 6 & 554 & 7942&0.069 \\
        A7  & 3 & 0.25 & 2.3 & 8 & 217 & 13093 &0.016\\
        A8 & 3 & 1    & 2.3 & 3 & 349 &3572 &0.097 \\
        A9 & 3 & 0.05 & 2.3 & 3 & 271 &1657 &0.163 \\ 
        A10  & 5 & 0.25 & 2.3 & 2 & 166 &506 &0.328 \\		
        A11  & 5 & 0.25 & 2.3 & 3 & 434 &1487 &0.291\\
        A12  & 5 & 0.25 & 2.3 & 8 & 828 & 6224 &0.133 \\
        A13  & 5 & 0.25 & 2.3 & 12 & 530 & >13200 &0.040 \\
        A14  & 5 & 0.25 & 2.3 & 16 & 0 & >13200 &0 \\

		\hline
	\end{tabular}
	\caption{Parameters of the various  $N$-body models with different initial half-mass radii ranging from $r_\mathrm{h,i}=$1 to 5 pc, different Galactocentric radii ranging from  $R_\mathrm{G}=$2 to 16 kpc, and metallicity of $Z=$ (0.05, 0.25, 1) $ Z_\odot $. The table also lists the name and main defining property of each model,  the DSC lifetime (column 6), the cluster lifetime (column 7), and their ratio (column 8). The canonical IMF ($\alpha_3=2.3$) is used in all models of set A.}
	\label{tab:initial_conditions}
\end{table}

\begin{table}
	\centering
	\begin{tabular}{cccccccc}        
        \hline
		  & &  & & Set B &  & &\\ 
		\hline  
		Model & $ r_\mathrm{h,i}\ $ & $ Z\ $ & $ \alpha_3 $ & $ R_\mathrm{G}\ $ & $\tau_\mathrm{DSC}$ & $\tau_\mathrm{cluster}$ & ${\tau_\mathrm{DSC}/}$ \\ 
		& $(\pc)$ & $(Z_\odot)$ & & $(\kpc)$ & $(\Myr)$& $(\Myr)$ &  ${\tau_\mathrm{cluster}}$
		\\
		\hline 
        B1  & 3 & 0.25 & 2.0  & 2 &358 & 598&0.598\\
        B2  & 3 & 0.25 & 2.0  & 3 &820 & 1326 & 0.618\\
        B3  & 3 & 0.25 & 2.0  & 4 &1145 & 1896 &0.603 \\
        B4  & 3 & 0.25 & 2.0  & 6 &1442 & 2592 & 0.556\\
        B5  & 3 & 0.25 & 2.0  & 8 & 1909& 3438 &0.555 \\
        B6  & 3 & 0.25 & 2.0  & 12 &2301 & 4359 &0.527\\
        B7  & 3 & 0.25 & 2.0  & 16 &2678 & 5444 &0.491\\
        B8  & 3 & 0.25 & 2.0  & 30 &4522 & 9400 &0.481\\
        B9  & 3 & 0.25 & 2.0  & 50 &5164 & >13200 &0.391\\
        B10 & 3 & 0.25 & 2.0  & 70 &2293 & >13200 &0.173\\
        B11  & 5 & 0.25 & 2.0  & 3 &770 & 1137 &0.677\\
        B12  & 5 & 0.25 & 2.0  & 6 &1417 & 2315 &0.612\\
        B13  & 5 & 0.25 & 2.0  & 8 &2095 & 3251 &0.644\\
        B14  & 5 & 0.25 & 2.0  & 12 &2485 & 4296 &0.578\\
        B15 & 5 & 0.25 & 2.0  & 16 &2994 & 5150 &0.581\\
        B16 & 5 & 0.25 & 2.0  & 30 &5048 & 9230 &0.547\\
        
        \hline
		  & &  & & Set C &  & &\\ 
		\hline 
	
        C1 & 1 & 0.25 & 1.7  & 3 & 959&  1012 &0.947\\
        C2 & 1 & 0.25 & 1.7  & 6 & 2132&  2185 &0.975\\
        C3 & 1 & 0.25 & 1.7  & 16 & 4042&  4095 &0.987\\
        C4 & 1 & 0.25 & 1.7  & 30 & 6459&  6512 &0.991\\
        C5 & 3 & 0.25 & 1.7  & 2 & 310&  362 &0.856\\
        C6 & 3 & 0.25 & 1.7  & 3 & 880&  950 &0.926\\
        C7 & 3 & 0.25 & 1.7  & 4 & 1408&  1481 &0.950\\
        C8 & 3 & 0.25 & 1.7  & 6 & 2066&  2145 &0.963\\
        C9 & 3 & 0.25 & 1.7  & 8 & 2238&  2335 &0.958\\
        C10 & 3 & 0.25 & 1.7  & 12 & 3484& 3595 & 0.969\\
        C11 & 3 & 0.25 & 1.7  & 16 & 4108& 4213 &0.975\\
        C12 & 3 & 0.25 & 1.7  & 30 & 6772& 6877 &0.984\\
        C13 & 3 & 0.25 & 1.7  & 50 & 10603& 10704 &0.990\\
        C14 & 3 & 0.25 & 1.7  & 70 & 13085& >13200 &0.991\\
        C15 & 3 & 0.25 & 1.7  & 100 & 13098&>13200 &0.992\\
        C16 & 3 & 0.25 & 1.7  & $\infty$ & 13116& >13200 &0.993\\
        C17 & 5 & 0.25 & 1.7  & 3 & 250&  294 &0.850\\
        C18 & 5 & 0.25 & 1.7  & 6 & 1826&  1922 &0.950\\
        C19 & 5 & 0.25 & 1.7  & 16 & 3402&  3576 &0.951\\ 
        C20 & 5 & 0.25 & 1.7  & 30 & 5533&  5708 &0.969\\
      
		\hline
	\end{tabular}
	\caption{The same as Table \ref{tab:initial_conditions} but for models with a top-heavy IMF ($\alpha_3 = 2.0,\ 1.7$).}
	\label{tab:initial_conditions_top_heavy}
\end{table}

In order to quantify the effect of a top-heavy IMF on the formation and evolution of DSCs, we use the collisional \Nbody code "NBODY7" \citep{Aarseth2012} which is an immediate descendant of the widely used NBODY6 direct \Nbody evolution code \citep{Aarseth2003, nbody7}. NBODY6/7 uses a fourth-order Hermite integration scheme, an individual time step algorithm to follow the orbits of cluster members, and invokes regularization schemes to deal with the internal evolution of small-N subsystems. In addition, NBODY6 treats single and binary stellar evolution in a comprehensive way from the zero-age main sequence (ZAMS) until their remnant phases, incorporating the SSE/BSE routines and analytical fitting functions developed by \citet{hurley2000}. NBODY7 incorporates important updates on two aspects of the single stellar evolution process. Firstly, compact object masses are assigned according to the prescription from \citet{Belczynski2008}. Secondly, the code implements the model presented by \citet{vink2001} to account for mass loss due to stellar winds, as described in \citet{Belczynski2010}. The high precision and efficient recipes for dealing with strong gravitational encounters, such as binary stars, are implemented in NBODY7 through the Algorithmic Regularization Chain of \citet{mikkola1999} instead of the classic Chain Regularization in NBODY6 \citep{Mikkola1993}. By utilizing this improved algorithmic regularization method for compact subsystems, it becomes possible to achieve general relativistic treatment through post-Newtonian terms and realistic parameters. Furthermore, this method provides a more thorough and reliable treatment of dynamically forming multiple systems in dense environments, particularly those involving massive objects like BHs.

A grid of 50 modeled star clusters is simulated to explore the effect of $r_\mathrm{h,i}$, $R_\mathrm{G}$, $Z$, and the stellar IMF on the start time and duration of the DSC phase. Initial stellar masses are distributed in the mass range of 0.07-150 $\Msun$, following a 3-segment power-law function as the IMF: 

\begin{equation}\label{eq:top_heavy_imf}
    \xi(m) \propto m^{-\alpha}: 
	\begin{cases}
		\alpha_1 = 1.3 & 0.07<\frac{m}{\Msun}<0.5    \\
		\alpha_2 = 2.3 & 0.50<\frac{m}{\Msun}<1.0 \\
		\alpha_3 & 1.00<\frac{m}{\Msun}<150
	\end{cases}\\
\end{equation} 
To cover the canonical \citep{kroupa2001} and top-heavy IMF \citep{MarksMichael2012, Kroupa2013}, $\alpha_3$ is varied from 2.3 (canonical Salpeter value) to 1.7. We perform three sets of simulations, one with a canonical IMF (A: $\alpha_3=2.3$), and other sets with top-heavy IMFs (B: $\alpha_3=2.0$ and C: $\alpha_3=1.7$). An overview of the performed simulations is given in \tabref{tab:initial_conditions} and \tabref{tab:initial_conditions_top_heavy}.

We performed simulations with an initial cluster mass of  $3\times10^4 \Msun$ and let the models evolve for $ 13.2 \Gyr $. The dissolution time is defined to be the time when the number of stars in the cluster declines to 10 stars. The initial positions and velocities of the stars in the cluster are chosen according to a Plummer phase-space distribution function \citep{plummer1911, Aarseth1974, Kroupa2008} in virial equilibrium. Initially, the models are not mass-segregated and do not include primordial binaries. However, all types of binaries and higher multiplicity systems are allowed to form during the evolution. Some of these dynamically formed binaries are retained in the clusters over their entire evolution.

It should be noted that the BH natal kicks are, to date, poorly constrained and understood from both observational and theoretical points of view (\secref{sec:Intro}). A common model \citep{Belczynski2008, Fryer2012} for supernova natal kick magnitude assumes NS-like kicks \citep{Hobbs2005} for BHs as well, but which are scaled down linearly with an increasing material fallback fraction, the so-called  canonical supernova kicks, upon which mass fallback typically results in about half of BHs being ejected at birth \citep{Chatterjee2017}. However, in this paper, we assumed the extreme limit for the retained BHs in modeled clusters (i.e. full retention fraction of BHs), to examine the pure effect of the initial BH mass fraction on the evolution and formation of the DSC phase. Therefore, the natal kicks at the time of the stellar remnant formation of NSs and BHs are assumed to be negligible, allowing all of them to remain in the cluster. A detailed discussion of the retention of stellar remnants in star clusters and ultra-compact dwarf galaxies is available in \citet{Jerabkova2017} and \citet{Pavlik2018}.

The initial velocity of the model star clusters is set for them to move on a circular orbit through the host Galaxy which is made up of three components: a central bulge, a disc, and a phantom dark matter halo potential \citep{Lughausen2015} that is scaled so the circular velocity at 8.5 kpc is 220 km/s. The bulge is modeled as a central point mass with a mass of $1.5\times10^{10}\Msun$. The gravitational potential of the disc is represented by the \citet{Miyamoto1975} profile,
\begin{equation}\label{eq:miyamoto}
	\phid = {\frac{-G\Md}{\sqrt{x^2 + y^2 + \left(a+\sqrt{b^2 + z^2}\right)^2}} }
\end{equation}
we used values of $a = 4 \kpc$ (disc scale length) and $b = 0.5 \kpc$ (Galactic thickness), while for the disc mass, we adopted $\Md = 5\times10^{10}\Msun$, as suggested by \citet{Xue2008}. We adopt a logarithmic potential for the dark matter halo of the MW,
\begin{equation}\label{eq:phi_halo}
    \phih  = {\frac{1}{2} \Vinfty^{2} \ln\left(\Rc^2 + R^2\right)},  	
\end{equation}
here, $R$ being the distance from the Galactic centre. The constant $\Rc$ is chosen such that the combined potential of the three components yields a circular velocity of $\Vinfty=220\kms$ in the disc plane at a distance of $8.5\kpc$ from the Galactic centre. 

All simulations were carried out on desktop workstations with Nvidia 1080 Graphics Processing Units at the Institute for Advanced Studies in Basic Sciences (IASBS).


\section{Dark Star Cluster}\label{sec:WH-DSC}

\begin{figure*}

  \centering\includegraphics[scale=1.0]{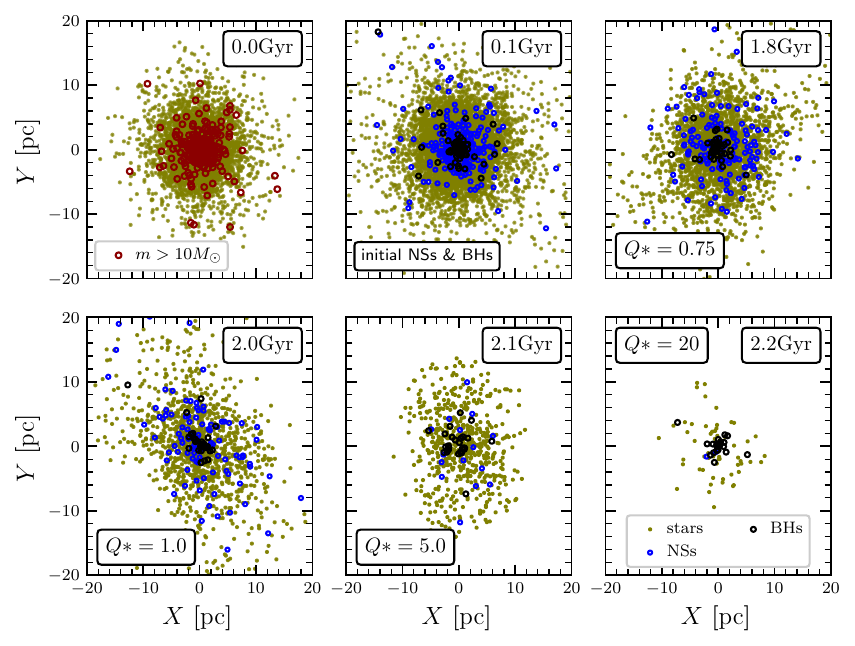}
  \caption{
  Top-down view of  model A4. The snapshots are depicted in a frame aligned with Galactocentric coordinates. Upper left: the initial distribution of low-mass stars (dots) and massive stars ($>10\Msun$) that end up as NSs or BHs (circles). Upper centre: the distribution of NSs (blue circles) and BHs (black circles) at $t=100\Myr$. Upper right: the cluster starts to exceed the virial equilibrium. Lower left: the beginning of the DSC phase. Lower centre: the cluster is significantly super-virial with $Q*=5.0$. Lower right: the final stage of the cluster's evolution, where the number of BHs is comparable to the luminous star population.}
  \label{fig:shots}
\end{figure*}

\begin{figure}

    \includegraphics[scale=0.505]{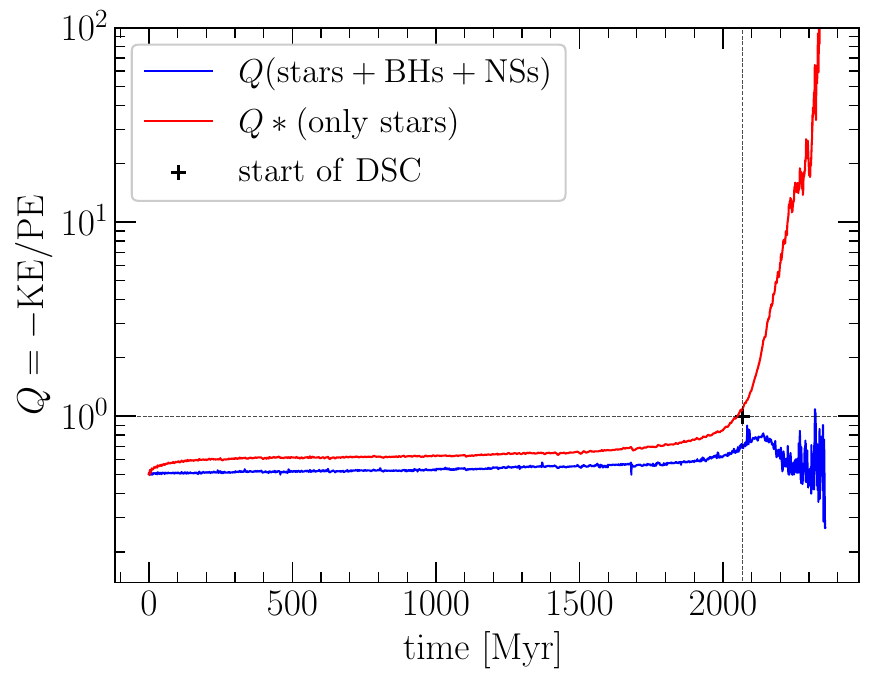}
    \includegraphics[scale=0.505]{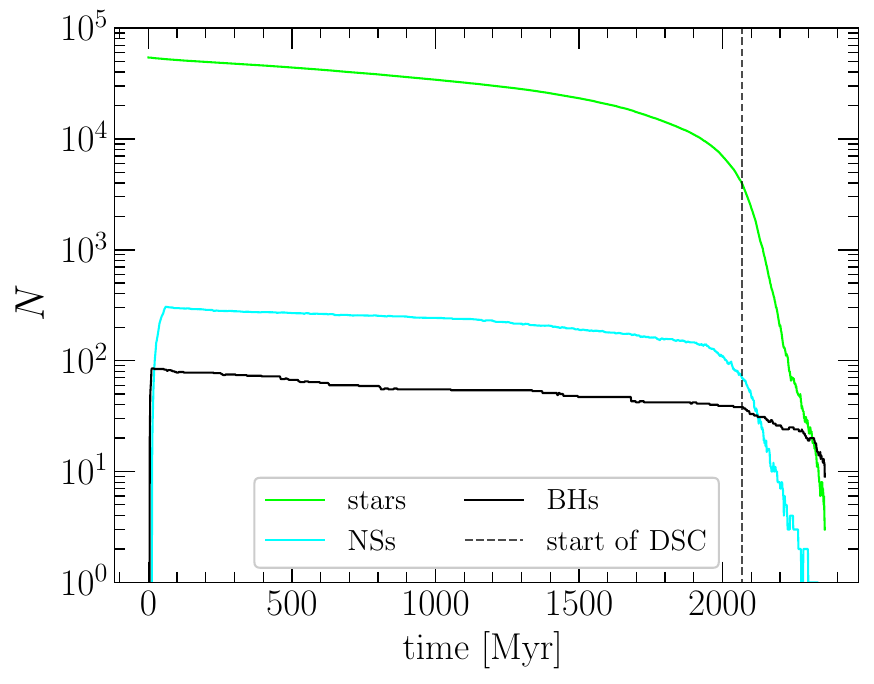}
    \includegraphics[scale=0.505]{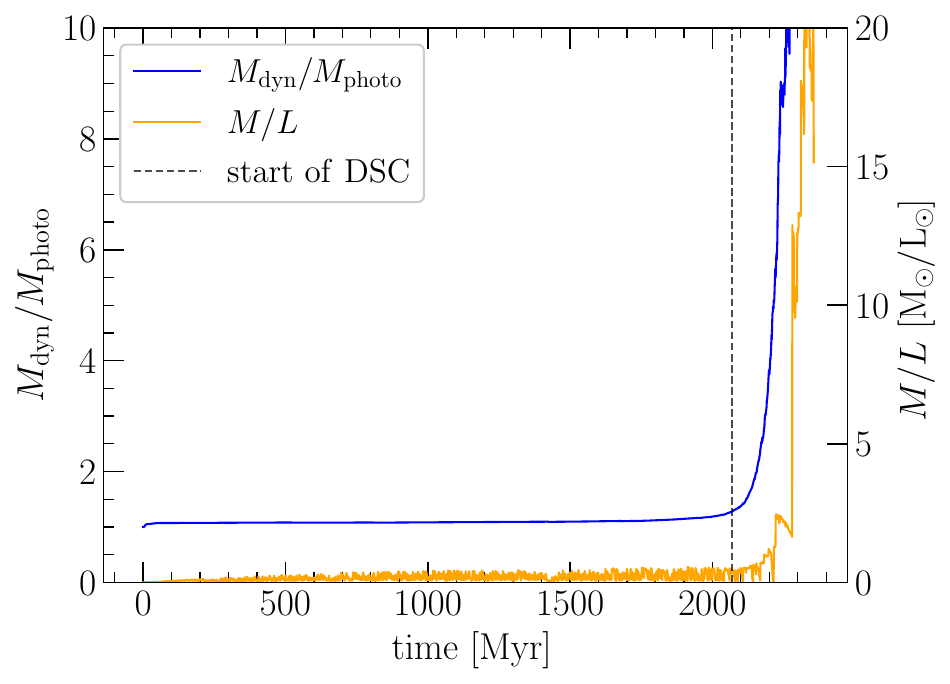}
    \caption{A typical example of the evolution of the virial coefficient of model A4. Top: the virial coefficient corresponding to the luminous stars only (red line) and the whole cluster (blue line). 
    Middle: variation of the number of luminous stars (lime line), neutron stars (cyan line), and black holes (black line). Bottom: time-evolution of dynamical mass to photometric mass ratio and dynamical mass-to-light ratio.}
    \label{fig:dsc}
\end{figure}

BHs in star clusters are formed in the first 10$\Myr$ of the cluster's evolution with masses of about 10-100 times larger than the average stellar mass of the cluster \citep{Belczynski2008, Belczynski2010}. If a large number of BHs remain in the cluster, they cannot reach energy equipartition with the low-mass stars and hence undergo runaway segregation toward the centre of the cluster. The dynamical friction on the dense stellar background segregates the BHs to the cluster center \citep{Spitzer1987} leading to the formation of a BHSub in the central part of the cluster. This process, which is known as Spitzer’s mass stratification instability, precludes thermal equilibrium. The Spitzer instability criterion is $\left(M_2/M_1\right)\left(m_2/m_1\right)^{3/2} > 0.16$, where $m_1$ and $m_2$ are the average mass of light and heavy stars, respectively; and $M_1$ and $M_2$ denote the total mass contained in the light and heavy components, respectively. The BHs segregation towards the central part of the cluster occurs over the following timescale:
\begin{equation}\label{eq:mass_segregation}
    t_\mathrm{seg}= \frac{\langle m \rangle}{\langle m_\mathrm{BH} \rangle} t_\mathrm{cc},  
\end{equation}
 where $t_\mathrm{cc}$ is the core-collapse timescale, $\langle m_\mathrm{BH} \rangle$  and $\langle m \rangle$ are the average mass of BHs and all stars, respectively. The value of $t_\mathrm{cc}$ depends on the initial half-mass relaxation time and is estimated to be about $t_\mathrm{cc}=15t_\mathrm{rh}$ for a single mass system and $t_\mathrm{cc}=0.2t_\mathrm{rh}$ for a realistic mass spectrum \citep{Heggie2003, Baumgardt2003, Fujii2014}. The centrally segregated BHSub is dynamically very active, and many BH-BH binaries (BBHs) form via the three-body interactions in the dense stellar environment \citep{Spitzer1987, Heggie2003}. Indeed, BHs of different masses can interact in a single close encounter and share their kinetic energy (KE). As a result, the massive BHs in the interaction make a binary BH, while the less massive BH picks up the excess KE released in the encounter and ejects to a higher (less bound) orbit. 
 
Through the subsequent encounters between BBHs and single BHs, the binary system becomes tighter, while the single BHs are scattered to higher orbits injecting their newly gained KE to the whole cluster via two-body interactions \citep{Banerjee2017}, which leads to an expansion
of the cluster. The scattered BHs eventually sink back to the cluster centre due to dynamical friction. With each encounter, the BBHs become more tightly bound and gain more recoil velocity. If the BBHs are sufficiently tight, during the next encounter, a significant amount of KE can be transferred to either the BBHs or single BH, leading to ejection from the cluster. This process is responsible for the self-depletion of the BH population from the clusters.

The self-depletion timescale depends on the number of BHs in the core. A cluster that retains a larger fraction of BHs will require more time for the dynamical self-emission of BHs compared to a similar cluster with fewer BHs. In other words, the dynamical self-ejection of BHs is a self-regulating process in which the BH ejection times are prolonged for more massive and numerous BHs \citep{Banerjee2017}. In star clusters orbiting in the inner part of the Galaxy, the outer skirt of star clusters strips rapidly due to the strong tidal field. When the removal timescale of low-mass stars from the outer region of the cluster due to the stronger tidal field is shorter than the self-depletion timescale of its BHSub, a new kind of star cluster (i.e., DSCs) which is dominated by BHs can be formed.

This evolutionary phase of star clusters is predicted for the first time by \citet{banerjee2011} using numerical simulations. During this phase, the cluster includes a few low-mass stars orbiting the central BHSub that observationally appear to be super-virial with a high $M_\mathrm{dyn}/L$ ratio. This is due to the velocity dispersion of the remaining stars being enhanced by the unseen BHs. The starting time of this phase can be defined as the time at which observable luminous stars (LSs), i.e., the nuclear-burning stars and the WDs, appear to be unbound or significantly super-virial. In other words, the DSC phase starts when the virial coefficient ($Q=-KE/PE$, where KE and PE are the kinetic and potential energy of the cluster) of the LSs, $Q*$, becomes greater than 1 \citep{banerjee2011}. Therefore, we use the criterion of $Q*>1$ to determine the starting time of the DSC phase.  It should be noted that all stars within twice the tidal radius are considered when calculating $Q$.

As an example, \figref{fig:shots} shows some projected snapshots of a modeled cluster (model A4) that enters into the DSC phase at about $2\Gyr$. Stars with masses $>10\Msun$ that evolve to NSs and BHs \citep{heger2003} are depicted in red at $t=0$. In the next snapshots, as the cluster evolves, BHs (black circles) and NSs (blue circles) are formed and segregate to the centre of the cluster, while many of the low-mass stars evaporate. Due to the BH and NS segregation and frequent encounters between BBHs and BHs, the released KE injects into the LSs and causes the whole cluster to expand. Therefore, the KE of the luminous low-mass stars increases and exceeds the self-equilibrium condition ($Q*>0.5$). When stars in the outer region evaporate, $Q*$ rises and the luminous sub-system begins to become super-virial, where the first signs of the DSC phase appear at $t=2.0\Gyr$ ($Q* \geq 1.0$). The following snapshots show the evolution of the DSC, where the LSs and even the NSs are stripped by the external tidal field, but the BHSub survives to the final stages of cluster dissolution. Note that the spherically asymmetric extensions of the cluster seen in \figref{fig:shots}, especially in the panels at $t=1.8 \ \mathrm{and} \ 2 \Gyr$, are the tidal tails. The presence of the BHSub elevates the background star density by accelerating the evaporation rate of LSs. Therefore, one of the major applications of the DSC phase is the enhancement of the tidal tail/stellar stream density \citep{Gieles2021}.

\figref{fig:dsc} displays the time-evolution of various dynamical parameters of model cluster A4, that can be used to introduce the DSC phase. The top panel shows a comparison of the virial-coefficient of the whole cluster's members ($Q$, blue line) enclosed by the tidal radius and the virial-coefficient of LSs ($Q*$, red line). As expected for self-gravitational systems, $Q$ remains constant at around $Q=0.5$ \citep{Heggie2003}, while  $Q*$ rises above $1.0$ when a significant fraction of LSs are tidally removed. The middle panel of \figref{fig:dsc}  illustrates that, at the onset of the DSC phase, approximately 90 per cent of LSs have escaped the cluster, while only half of the BHs have been ejected. Furthermore, it highlights that NSs evaporate earlier than BHs. Therefore, in the final stages of the cluster evolution, where the potential of the dark sub-system gradually becomes important, the cluster shows its DSC phase that is dominated by BHs. The ratio of the dynamical mass to photometric mass ($M_\mathrm{dyn}/M_\mathrm{photo}$) is another parameter that can indicate the DSC phase, where $M_\mathrm{photo}$ is measured by summing the masses of all luminous objects (consisting of nuclear-burning stars and WDs). The bottom panel of \figref{fig:dsc} shows that both $M_\mathrm{dyn}/M_\mathrm{photo}$ and $M_\mathrm{dyn}/L$ remain almost constant until the final stage of the evolution, increasing dramatically around the starting time of the DSC phase.

Several BBHs (and other types of binaries) form during the lifetime of our model clusters. Of these, 54 BH-BH, 3 BH-LS, 1 BH-NS, and 1 NS-LS were formed in the A4 model. Our simulations have shown that only a few BH mergers occur, all of which happened in the first 500$\Myr$. The number of BH mergers in GCs generally increases linearly with the cluster mass \citep{Kremer2020}. However, for GCs with top-heavy IMFs, the number of BH mergers increases super-linearly with mass \citep{Weatherford2021}. So, GCs of a typical mass should actually be significant BH merger sources, especially GCs born with top-heavy IMFs. However, the rate of BBH mergers is more intense in the early stages of cluster evolution and substantially decreases over time, especially for GCs with canonical IMFs \citep{Banerjee2010, Abadie2011,Weatherford2021}. This means that if the cluster evolves into a DSC during the later stages of its evolution, it is unlikely to detect BBH mergers that produce detectable gravitational waves during the DSC phase. Nevertheless, the remnant sub-system in DSCs can be a rich source for X-ray binaries and soft gravitational waves due to the large number of binaries with an accretor component, i.e., a BH or NS \citep{Tauris2006}.

The characteristics of the cluster that would evolve into the DSC phase can be determined as follows: 
\begin{itemize}

\item The evaporation time of LSs ought to be shorter than the BHSub self-depletion timescale.

\item LSs go out of equilibrium and become super-virial. As a consequence, $Q*$ experiences a substantial increase, sometimes surpassing 100.

\item Accelerating the evaporation rate of LSs enhances tidal tail/stellar stream density.

\item The dynamical mass-to-light ratio of the cluster increases during the DSC phase, almost like $Q*$.

\item The ratio of BHSub mass to the total mass of the cluster,

\begin{equation}
    \widetilde{\mathit{M}}_\mathrm{BH}(t)= M_\mathrm{BH}(t)/M(t),
\end{equation} \label{MBH}
increases over time, and the value of $M_\mathrm{dyn}/M_\mathrm{photo}$ grows strongly similar to $Q*$. At $Q*=1$, $M_\mathrm{dyn}/M_\mathrm{photo}$ will be approximately equal to 1.4, according to the average value of our simulated model results.

\item The presence of a BHSub in a cluster can lead to the formation of a significant number of binaries containing BHs and BBHs, which can serve as sources for X-ray binaries and soft gravitational waves.  If the initial cluster is very dense, BBH merging can happen, which is the source of observable gravitational waves. We expect this event to occur early in the evolution of the cluster. However, if the cluster evolves into the DSC phase at the late stages of its evolution, we would not expect to detect a gravitational wave signal in the DSC phase.
\end{itemize}

In this paper, we define the scaled DSC lifetime as follows:
\begin{equation}\label{eq:dsc_power}
    \widetilde{\mathit{\tau}}_\mathrm{DSC} = \frac{\tau_\mathrm{DSC}}{\tau_\mathrm{cluster}},  
\end{equation}
where $\tau_\mathrm{DSC}$ is the time interval during which the cluster is in its DSC phase and $\tau_\mathrm{cluster}$ is the total lifetime of the cluster. The value of  $\widetilde{\mathit{\tau}}_\mathrm{DSC}$ determines how much of the cluster's lifetime is in the DSC phase. A zero value of $\widetilde{\mathit{\tau}}_\mathrm{DSC}$ implies that the cluster will not reach the DSC phase at all. On the other hand, the closer the value of $\widetilde{\mathit{\tau}}_\mathrm{DSC}$ is to 1, the more the cluster spends its lifetime in the DSC phase. Therefore,  $\widetilde{\mathit{\tau}}_\mathrm{DSC}$ is a good parameter to determine how much the BHSub dominates the cluster evolution. To assess the strength of BHSub dominance, one can calculate the difference between the self-depletion time of the  BHSub and the evaporation time of LSs, divided by the BHSub self-depletion time ($(\tau_\mathrm{dep}-\tau_\mathrm{eva})/\tau_\mathrm{dep}$). A negative value of this term indicates that the cluster will not evolve to the DSC phase. This term is directly proportional to $\widetilde{\mathit{\tau}}_\mathrm{DSC}$, which means that as the value of this term increases, the value of $\widetilde{\mathit{\tau}}_\mathrm{DSC}$ will also increase.


\section{Results}\label{sec:result}

In order to study the influence of a top-heavy IMF on the formation and  evolution of the DSC phase in detail, we calculated two sets of models, one in which the IMF is canonical  ($\alpha_3=2.3$), and one set with an initially top-heavy IMF ($\alpha_3=2.0$ and $1.7$). Both will be discussed in the following. The IMF of models in the following section (\secref{sec:Canonical}) are canonical. Then we evaluate the effect of the top-heaviness of the IMF on the results (\secref{sec:top}). In \secref{sec:DISCUSSION}, we combined the results from the previous two sections and identified the transition boundary to the DSC phase for various levels of top-heaviness.


\subsection{Canonical IMF}\label{sec:Canonical}

Before we start looking into the effect of top-heavy IMF, we examine how changing  $r_\mathrm{h,i}$, $Z$ and $R_\mathrm{G}$ of a star cluster with a canonical IMF influences the formation and evolution of the DSC phase. This will allow us to make inferences about the sensitivity of the results on choosing these crucial initial parameters.

\subsubsection{The impact of the initial half-mass radii on the formation of the DSC phase}\label{sec:eff Rh}

\begin{figure*}
    
    \centering\includegraphics[scale=0.97]{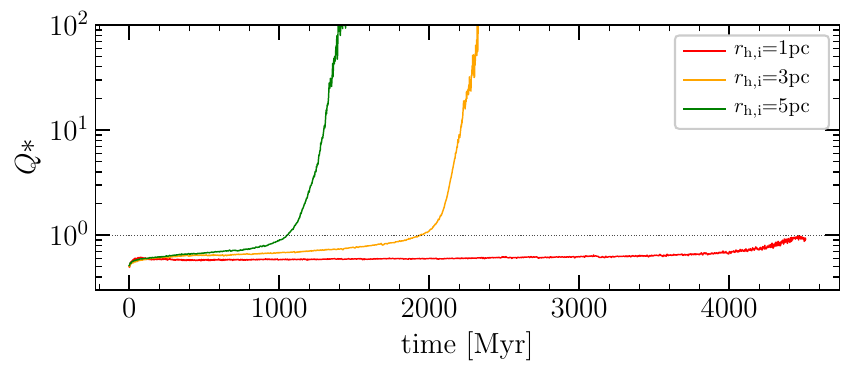}
    \centering\includegraphics[scale=1.0]{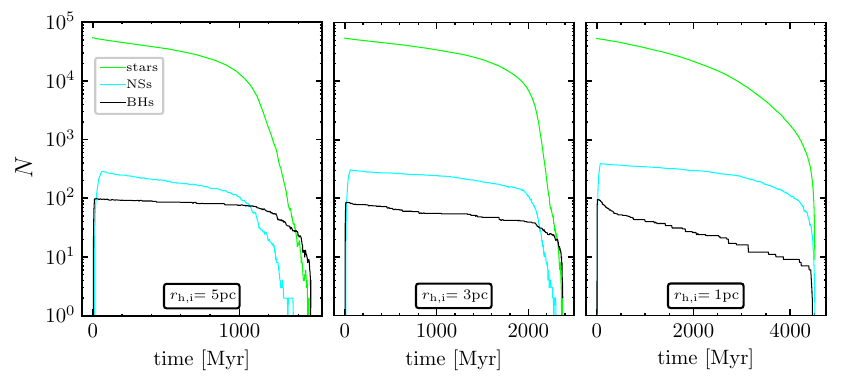} 
    \caption{Top: time variation of $Q_*$ for clusters with three different half-mass radii of $r_\mathrm{h,i}$ = 1, 3, and 5 pc (red, orange, and green lines, respectively). Bottom: the number of stars (lime line), neutron stars (cyan line), and black holes (black line) as a function of time for clusters depicted in the top panel.}
    \label{fig:size}
\end{figure*}

\begin{figure}

  \centering\includegraphics[scale=0.505]{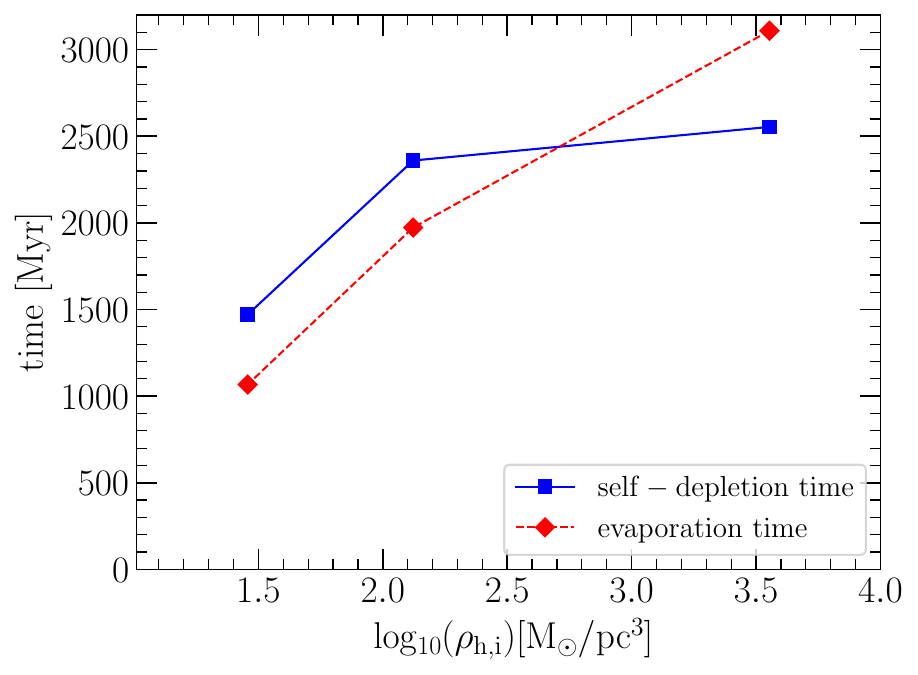}
  \caption{For simulations in set A with $R_\mathrm{G}=3\kpc$, the blue solid and red dashed lines represent the self-depletion and evaporation times, respectively, of BHSub and LSs.}
  \label{fig:dep-eva-size}

\end{figure}

Using \Nbody models of clusters with BHs, \citet{Gieles2023} showed that the initial density is a critical parameter in setting the dynamical retention of BHs such that the mass-loss rate of the BH population due to dynamical ejections before the cluster fills the tidal radius only depends on the square root of the initial density. In this study, we aim to investigate how the initial density of star clusters affects the formation and evolution of DSCs. Since all the modeled clusters have the same initial mass of $3\times10^4 \Msun$, varying the value of $r_\mathrm{h,i}$ will result in different initial densities within the half-mass radii, $\rho_\mathrm{h,i} = 3 M_{\mathrm{cluster}}/{8 \pi  r_\mathrm{h,i}^3}$.

We calculate models with different $r_\mathrm{h,i}$ of 1, 3, and 5 $\pc$,  moving on a circular orbit at $R_\mathrm{G}=3\kpc$. The evolution of the virial coefficient ($Q*$) of these clusters is shown in the top panel of \figref{fig:size}. The modeled cluster with $r_\mathrm{h,i}=5 \pc$ (A11) is not gravitationally bound enough to resist the escape of background stars that have acquired the energy generated in BHSub. Therefore, LSs evaporate rapidly and the cluster reaches the DSC phase at about $t=1\Gyr$. The model with $r_\mathrm{h,i}=3 \pc$ (A4) reaches the DSC phase later at about $t=2\Gyr$ and the model with  $r_\mathrm{h,i}=1 \pc$ (A2) does not experience the DSC phase before dissolution. The bottom panel of \figref{fig:size} shows the time evolution of the number of LSs (lime line), NSs (cyan line), and BHs (black line) for these clusters. As can be seen, the number of LSs decreases significantly in model A11 ($r_\mathrm{h,i}=5 \pc$), while the number of BHs remains almost constant until the final stage of the cluster evolution, when the cluster enters the DSC phase and the cluster becomes dominated by BHs. After the evaporation of LSs and NSs from the cluster, the BHSub starts to be tidally stripped.

The models with the smaller half-mass radii ($r_\mathrm{h,i}=3 \pc$ and $1\pc$ pc) are relatively more gravitationally bound, which leads to slower evaporation of LSs, while the rate of few-body encounter is higher within the BHSub, leading to a faster BH self-depletion. In other words, the energy generated from the BHSub is not sufficient to bring the LSs to the escape velocity. On the other hand, collisions in the BHSub are more frequent in clusters with higher densities, leading to faster BH self-depletion compared to the LS evaporation. As a result, the number of BHs decreases with a steeper slope compared to the cluster with $r_\mathrm{h,i}=5 \pc$ (\figref{fig:size}, bottom panel). In the case of $r_\mathrm{h,i}=1 \pc$, the cluster does not exhibit any sign of the DSC phase and LSs never become super-virial as the virial coefficient does not exceed $Q*=1$ (top panel of \figref{fig:size}, red curve). As shown in the bottom panel of \figref{fig:size}, while the number of BHs decreases with a relatively sharp slope, NSs start to replace them by accumulating in the central part as a result of the Spitzer instability.

In a nutshell, when BHs segregate due to dynamical friction, they give their initial KE to low-mass stars. In addition, during the super-elastic encounters between BBHs and single BHs, a certain amount of KE is imparted to background stars (\secref{sec:WH-DSC}). This amount of energy released from the BHSub can take LSs to escape velocity, thus increasing their evaporation rate, and the cluster eventually reaches the super-virial phase. In dense star clusters, the gravitational potential is deeper; hence, the released energy from the BHSub would not be enough to bring the LSs to escape velocity. So, the influence of the BHSub on the evaporation rate of LSs is more pronounced in low-density clusters compared to dense star clusters. On the other hand, the BHs are immersed in a deep central potential, before the evaporation of the LSs. During this period, the escape rate of BHs depends mostly on the few-body encounters between BBHs and single BHs rather than the tidal field of the host galaxy \citep{Breen2013, Longwang2020}. The LSs in the cluster halo shield the BHSub from the Galactic tidal force. Since few-body encounters in the BHSub are more frequent in a denser cluster, the BHSub will be depleted faster. As a consequence, in a dense cluster, the self-depletion time of the BHSub is shorter than the evaporation time of LSs and the cluster will not evolve into the DSC phase.

It is important to note some details about the relationship between the evaporation timescale of LSs and the initial density of clusters. When we try to determine whether the evaporation time of LSs increases or decreases with initial density, we need to consider two conflicting effects. Firstly, reducing the cluster density makes it easier for LSs to evaporate by lowering the cluster escape velocity. This can also be thought of as the cluster overflowing its tidal boundary. Secondly, lower density also slows down the evaporation of LSs by increasing the two-body relaxation timescale \citep{Spitzer1987, Heggie2003}. Which of these competing effects dominates depends on how tidally filling the cluster is at birth. However, the DSC phase is determined by the discrepancy between the timescales of LS evaporation and BHSub self-depletion, not by their individual rates. Indeed, if the opposite scaling is achieved (faster evaporation of LSs with higher density), dense clusters persistently fail to transition to the DSC phase. In such cases, while the evaporation time of LSs of the denser cluster becomes shorter than that of the low-density counterpart, the self-depletion time of their BHSub also contracts due to heightened encounters within BHSub, offsetting the faster evaporation time of LSs. The stronger gravitational binding of dense clusters restricts the energy injected by the BHSub to enhance the evaporation rate of their LSs. Therefore, in dense clusters, the evaporation rate of LSs is mainly determined by two-body relaxation. In clusters with a lower density, although the evaporation time is extended, the self-depletion time is also prolonged. Since these clusters have a lower gravitational binding, the BHSub can exert a dominant influence on the evaporation rate of LSs, which facilitates the transition to the DSC phase.

The evaporation and the BHSub self-depletion timescales are compared for clusters with different densities located at $R_\mathrm{G}=3\kpc$ in \figref{fig:dep-eva-size}. The cluster evaporation time and the BHSub depletion time are defined as the times when 85 per cent of the initial mass of LSs and BHs are ejected from the cluster, respectively. The DSC phase appears when the evaporation time (red line) of LSs is shorter than the BHSub self-depletion timescale (blue line). As the value of $\rho_\mathrm{h,i}$ increases, the depletion and evaporation times become closer to each other. This leads to a decrease in $\widetilde{\mathit{\tau}}_\mathrm{DSC}$, indicating that the cluster spends less time in the DSC phase. As expected, for higher density, the BHSub self-depletion timescale becomes faster than the LS evaporation timescale, meaning that the DSC phase does not appear.

For a cluster with a higher stellar concentration, we expect a faster depletion of the BHSub. However, as shown in \figref{fig:dep-eva-size}, for clusters located deep in the Galactic potential ($R_\mathrm{G}=3\kpc$), the self-depletion timescale of compact models (such as A2) is longer than that of extended clusters (A4 and A11). This is due to the tidal effect of the host Galaxy, which strips the BHSub after LS evaporation. Therefore, under isolated conditions, we would expect the BHSub of compact models to dissolve faster than extended clusters.

\subsubsection{The impact of the Galactocentric distance on the formation of the DSC phase}\label{sec:eff Rg}

\begin{figure}

    \includegraphics[scale=0.1818]{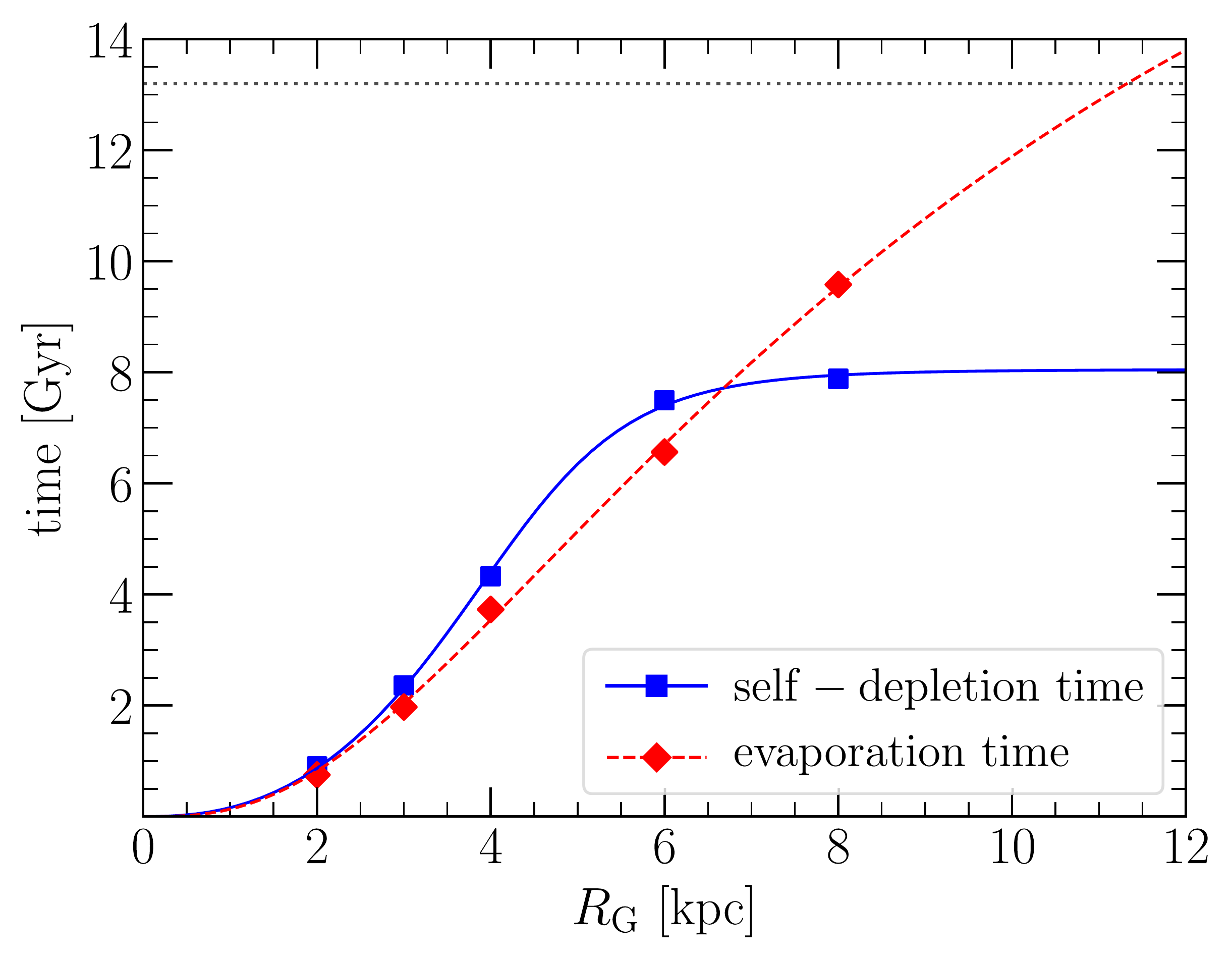}
    \includegraphics[scale=0.49]{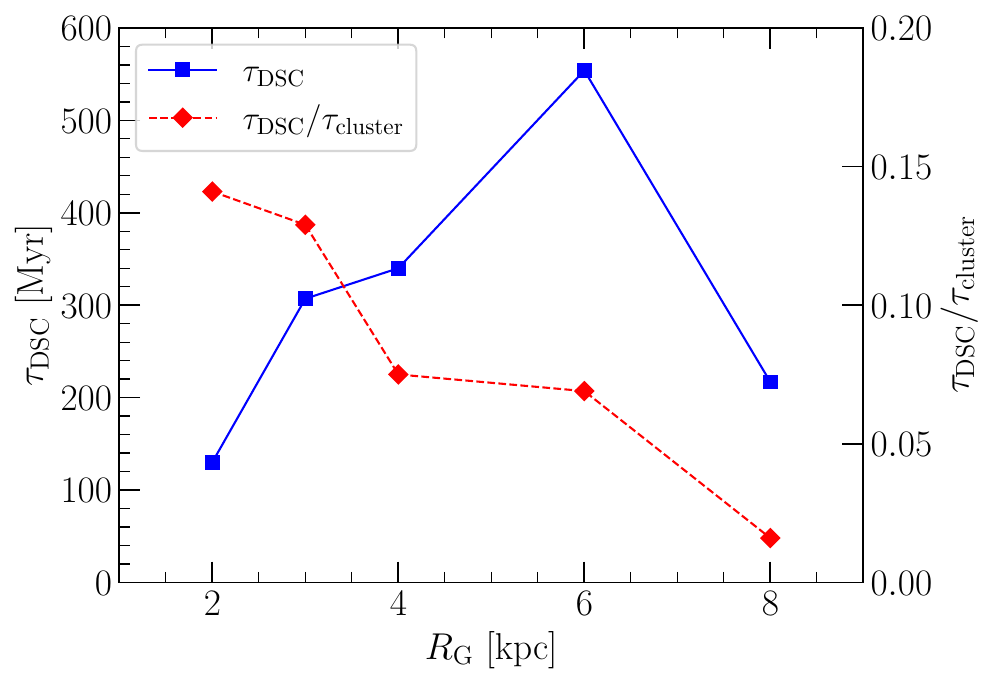}
    \caption{Top: evaporation time of LSs (red diamonds) and self-depletion time of the BHSub (blue squares) at different $R_\mathrm{G}$ for models A3 to A7. Fitted curves, represented by a blue solid line and a red dashed line, respectively, capture trends in self-depletion and evaporation time. The horizontal dashed line shows the Hubble time (13.2 $\Gyr$). Bottom: the DSC lifetime (the left axis) and the scaled DSC lifetime ($\widetilde{\mathit{\tau}}_\mathrm{DSC}$, Eq. \ref{eq:dsc_power}, the right axis) for different $R_\mathrm{G}$ are shown by blue solid and red dashed lines, respectively.} 
    \label{fig:rg}
\end{figure}

It is well known that by increasing the Galactocentric distance of an orbiting cluster, which means reducing the tidal field of the host galaxy, the evaporation timescale increases. Conversely, the escape rate of BHs from the BHSub (i.e., the self-depletion timescale of the BHSub) is predominantly controlled by the frequency of few-body encounters within the BHSub, rather than the tidal field, as long as the halo of low-mass stars has not evaporated. Consequently, for a tidally underfilling cluster, increasing $R_\mathrm{G}$ does not have a major effect on the self-depletion timescale of the BHSub \citep{Breen2013, Longwang2020}. Therefore, we expect that $R_\mathrm{G}$ plays an important role in the formation of the DSC phase. To investigate the effect of $R_\mathrm{G}$ on the formation of the DSC phase, we calculated models orbiting at different Galactocentric distances, $R_\mathrm{G}=2, 3, 4, 6 \ \mathrm{and} \ 8\kpc$ (models A3-A7), each starting with an initial half-mass radii of $r_\mathrm{h,i}=3\pc$.

As can be seen in \figref{fig:rg}, the evaporation time increases with $R_\mathrm{G}$ and becomes longer than the self-depletion time beyond 6 $\kpc$. Then the BHSub self-depletion time becomes independent of the tidal field and does not vary as $R_\mathrm{G}$ increases (it will eventually converge to about $8\Gyr$). The bottom panel of \figref{fig:rg} shows the lifetime of the DSC phase ($\tau_\mathrm{DSC}$) vs. $R_\mathrm{G}$. The  value of $\tau_\mathrm{DSC}$ rises up to 600$\Myr$ at around 6 $\kpc$. This is because after the evaporation of LSs, in a weak external field, it takes a longer time for a BHSub to deplete. As $R_\mathrm{G}$ increases to above 6 $\kpc$, the $\tau_\mathrm{DSC}$ decreases and eventually reaches zero, which means that the cluster does not enter the DSC phase. This is because the evaporation time increases with $R_\mathrm{G}$, while the self-depletion time remains almost constant. The $\widetilde{\mathit{\tau}}_\mathrm{DSC}$ (red line) in the bottom panel of \figref{fig:rg} describes the fraction of cluster lifetimes spent in the DSC phase. $\widetilde{\mathit{\tau}}_\mathrm{DSC}$ decreases with increasing $R_\mathrm{G}$ such that the cluster spends 14 per cent of its life in the DSC phase at $R_\mathrm{G}=2\kpc$, while it is 1 per cent at $R_\mathrm{G}=8\kpc$.

\subsubsection{The impact of the initial metallicity of the cluster on the formation of the DSC phase}\label{sec:eff Z}

\begin{figure}
\centering\includegraphics[scale=0.57]{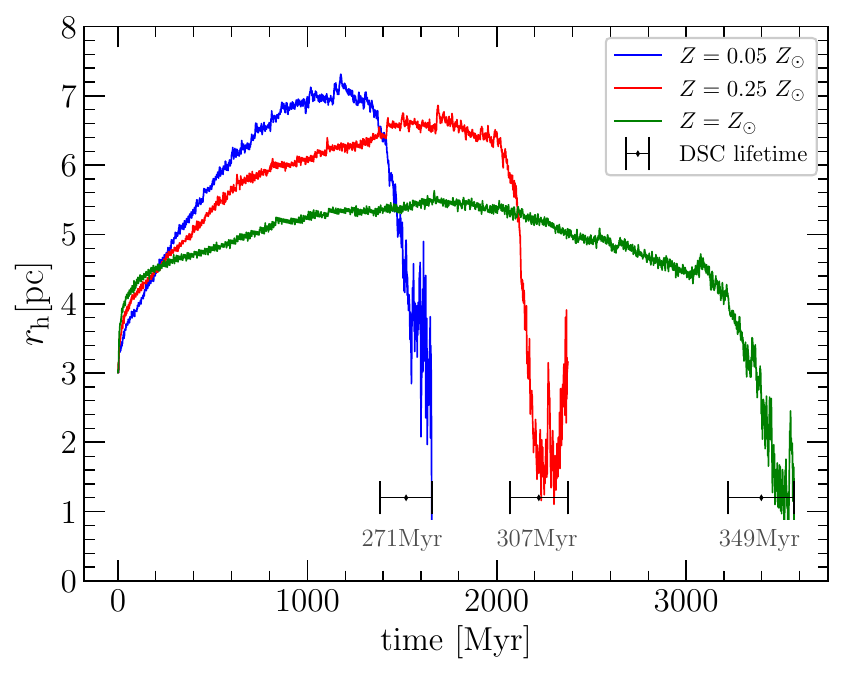}
  \caption{The evolution of half-mass radii for three clusters with three different metallicities of $Z = 0.001$ (blue line), $Z = 0.005$ (red line), and $Z = 0.02$ (green line). The time intervals ($\tau_\mathrm{DSC}$) during which the clusters are in their DSC phase are shown.}
  \label{fig:z}
\end{figure}

Metallicity in star clusters has a significant impact on their dynamic evolution. This is because it affects the evolution of massive stars, the rate at which they lose mass through stellar winds, and the final mass of the remnants \citep{trani2014}. Studies have shown that the retention mass fraction of BHs at birth is higher in metal-poor clusters than in metal-rich clusters \citep{Shanahan2015}. Moreover, in metal-poor clusters, much more massive BHs can form compared to metal-rich clusters \citep{vink2001,vink2005,Belczynski2010}. As a result, the average mass of BHs and their retained number increases as the metallicity of the cluster decreases. The energy generated from the segregation of BHs and their dynamical interactions increases with the increasing number of BHs and their average mass. Consequently, in metal-poor clusters, the energy injection from the BHSub is larger, leading to faster expansion and a higher evaporation rate of LSs. In addition, the larger number of BHs leads to a longer self-depletion time for the BHSub. Both of these factors lead metal-poor clusters to spend a larger fraction of their lifetime in the DSC phase.

To investigate the effect of $Z$ on the DSC phase, we consider three clusters with the same mass, $r_\mathrm{h,i}$, and $R_\mathrm{G}$, but different metallicities of 0.05$Z_{\sun}$, 0.25$Z_{\sun}$ and $Z_{\sun}$ (models A4, A8, and A9). The BHSub masses of these models were 1724$M_{\sun}$, 1226$M_{\sun}$, and 734$M_{\sun}$, respectively, at birth. The time evolution of the half-mass radii, $r_\mathrm{h}$, of these clusters is shown in \figref{fig:z}. During the first $\approx200\Myr$, the cluster with Solar metallicity expands faster than the sub-Solar metallicity clusters due to the dominant role of stellar evolution. Metal-rich clusters lose more mass through stellar winds and supernova expulsion during the early stages  \citep{vink2001, vink2005, Schulman2012, Mapelli2013}. However, after $\approx200\Myr$, this trend reverses, and the sub-solar clusters expand more rapidly due to more energy generation from segregated BHs. As a result, metal-poor clusters expand to a larger size and dissolve earlier than metal-rich clusters (consistent with \citealt{Debatri2022}). The $\mathit{\tau}_\mathrm{cluster}$ of the modelled cluster with $Z=0.05Z_\odot$ is about half of the modelled cluster with $Z=Z_\odot$, while $\widetilde{\mathit{\tau}}_\mathrm{DSC}$ is twice as large for $Z=0.05Z_\odot$. We also compared the mass fraction of BHs retained in these clusters when 75 per cent of their initial mass had been lost. Our findings indicate that when the metallicity is decreased from $Z=Z_\odot$ to $Z=0.05Z_\odot$, the mass fraction of BHs rises from $0.055$ to $0.115$ at a point when only 25 per cent of the initial mass is left in the clusters.  Summarising, these results thus show that the DSC phase is shorter at low metalicity but comprises a longer fraction of the cluster's lifetime than at high metalicity.

\subsubsection{The Combined effect of initial density and Galactocentric distance on the formation of the DSC phase }

\begin{figure}

  \centering\includegraphics[scale=0.43]{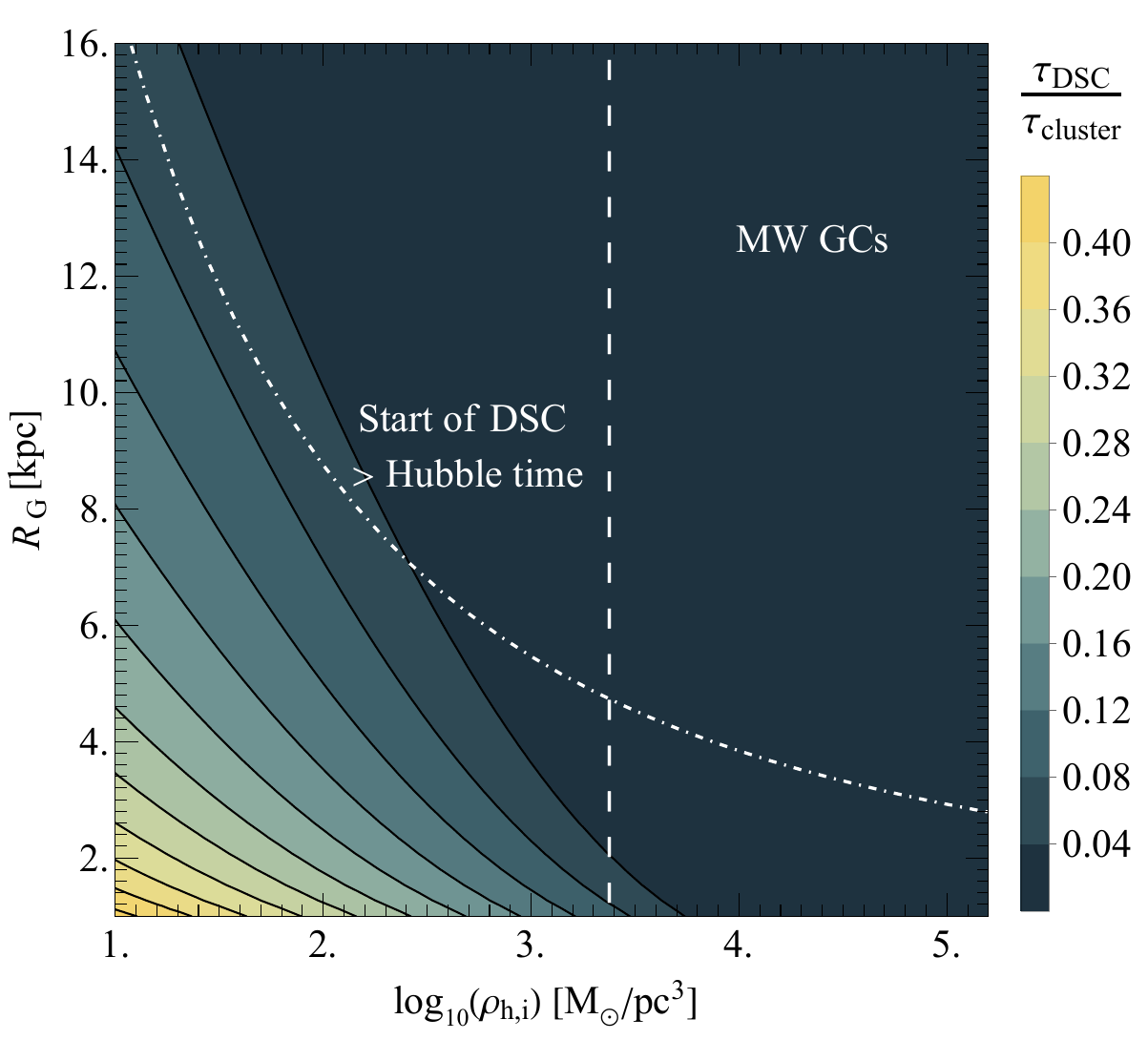}
  \caption{$\widetilde{\mathit{\tau}}_\mathrm{DSC}$ is shown as a function of $\log _{10}(\rho_\mathrm{h,i})$ and $R_\mathrm{G}$. The right side of the perpendicular with a dashed line shows an acceptable range for the initial density of the MW GCs. The dashed-dotted line represents the curve where the starting time of DSC equals the Hubble time.}
  \label{fig:Rg-Size-2,3}

\end{figure}

According to \tabref{tab:initial_conditions}, $\widetilde{\mathit{\tau}}_\mathrm{DSC}$ decreases for clusters constructed based on the canonical IMF by increasing $\rho_\mathrm{h,i}$ and $R_\mathrm{G}$. We derived the best-fitting function for $\widetilde{\mathit{\tau}}_\mathrm{DSC}$ in dependence of $\log _{10}(\rho_\mathrm{h,i})$ and $\log _{10}(R_\mathrm{G})$ of the modelled clusters in the following mathematical form:

\begin{equation}\label{eq:P2,3}
    \widetilde{\mathit{\tau}}_{\mathrm{DSC},\alpha_{3}=2.3}\left(R_\mathrm{G},\rho_\mathrm{h,i} \right)=a (\rho_\mathrm{h,i})  \log _{10}\left(\frac{R_\mathrm{G}}{\text{kpc}}\right)+b (\rho_\mathrm{h,i}),  
\end{equation}
where $a$ and $b$ are the best-fitting parameters which are themselves a function of $\log _{10}(\rho_\mathrm{h,i})$ as follow:
\begin{equation}\label{eq:a2,3}
    \left\{a,b\right\} (\rho_\mathrm{h,i}) =\left\{a_{1},b_{1}\right\}\log _{10}\left(\frac{\rho_\mathrm{h,i}}{\mathrm{M}_{\odot }\text{pc}^{-3}} \right)+\left\{a_{2},b_{2}\right\}. 
\end{equation}
\figref{fig:Rg-Size-2,3} illustrates $\widetilde{\mathit{\tau}}_\mathrm{DSC}$ as a function of $\log _{10}(\rho_\mathrm{h,i})$ and $R_\mathrm{G}$. $\widetilde{\mathit{\tau}}_\mathrm{DSC}$ increases as $R_\mathrm{G}$ and $\rho_\mathrm{h,i}$ decreases and reaches its maximum of 44 per cent for the lowest values that we adopted for $R_\mathrm{G}$ and $\rho_\mathrm{h,i}$. Star clusters with an initial density $\log _{10}(\rho_\mathrm{h,i})>3.7$ do not experience the DSC phase. The DSC phase within the Hubble time can only be observed in clusters with $R_\mathrm{G}<16\kpc$.

According to the \citet{Marks2012} relation, a cluster with a birth mass of $3\times10^4 \Msun$ has an initial half-mass radii of 0.38$\pc$. It's worth noting that our simulations do not account for gas expulsion, so we have to consider the starting conditions of our models as post-gas expulsion. \citet{Baumgardt2007} found that assuming a star formation efficiency of about $0.3$, massive clusters expand by a factor of $\simeq 3$ during the gas expulsion phase. This expansion factor is almost independent of the gas removal rate. By applying this expansion factor, the post-gas expulsion density of a cluster with a mass of $3\times10^4 \Msun$ will be about $\log _{10}(\rho_\mathrm{h,i})=3.4$. We adopt this value as a lower limit for the initial density of the MW GCs (vertical white dashed line in \figref{fig:Rg-Size-2,3}). Note that the initial density of the MW GCs is on average higher than what our model suggests. This is because the MW GCs tend to be more massive than the clusters we've used in our calculations.

Based on our analysis, it is unlikely for any MW GCs to enter the DSC phase if we consider the canonical IMF. This is because, in such clusters, the BHSub self-depletion happens at a faster rate than the evaporation of LSs. Over time, the ratio of BHSub mass to the total mass of the cluster ($\widetilde{\mathit{M}}_\mathrm{BH}(t)$, eq. \ref{MBH})  gradually decreases until it ultimately reaches zero. The white dashed-dotted curve separates the initial conditions of the star clusters into two groups: those that enter the DSC phase before the Hubble time (below the curve) and those that enter after (above the curve). This means that even if a cluster in the upper region could eventually evolve into the DSC phase, it would do so after the Hubble time.

\begin{figure}

  \centering\includegraphics[scale=0.43]{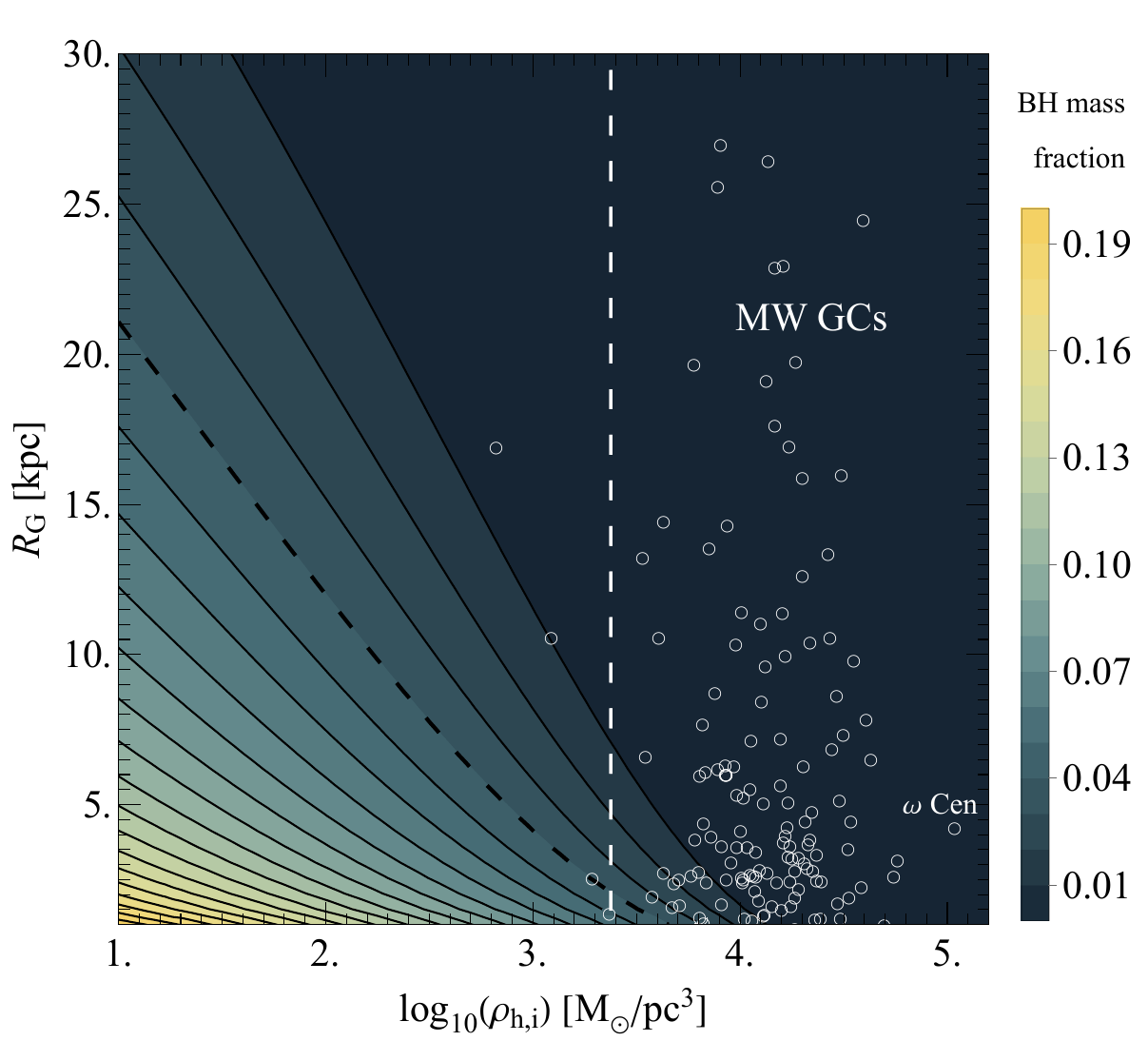}
  \caption{BH mass fraction of the modeled clusters with a canonical IMF at the moment when they have lost 75 per cent of their mass since formation.  The dashed contour represents a BH mass fraction of 0.04. The white circles indicate where the 166 MW GCs are located in the 2D space of initial density and Galactocentric distance. The vertical with dashed line is as in Fig. \ref{fig:Rg-Size-2,3}.}
  \label{fig:75}

\end{figure}

The MW GCs have on average lost nearly 75 per cent of their initial stellar masses through the stellar evolution or long-time dynamical evolution \citep{Webb2015,Baumgardt2017-75per,Baumgardt2019mean}. Therefore, the BH mass fraction of the modeled clusters is plotted in \figref{fig:75}, which shows the fraction of BHs when 75 per cent of the initial mass of the clusters has been lost. For comparison with the MW GCs, the mean Galactocentric distance \citep{Baumgardt2021,Vasiliev2021} and post-gas expulsion density of the 166 GCs are shown with white circles. To calculate the post-gas expulsion density of MW GCs, we consider 4 times their present-day dynamical mass (\citealt{Baumgardt2018}, additionally, we make use of the GC database developed by Baumgardt and Sollima\footnote{https://people.smp.uq.edu.au/HolgerBaumgardt/globular/}) as their initial stellar mass. Using the \citet{Marks2012} relation, we obtain the initial half-mass radii, which is multiplied by a factor of 3 to obtain the radius of star clusters after gas expulsion \citep{Baumgardt2007}. The black dashed curve corresponds to the mass fraction of BHs predicted by stellar evolution (birth fraction), which is approximately  $\widetilde{\mathit{M}}_\mathrm{BH}(0)=0.04$. For clusters below this curve, the BH fraction increases over time, while for dense clusters located at large $R_\mathrm{G}$ (above this curve), the BH fraction decreases over time until it reaches zero. This black dashed curve represents the boundary beyond which the cluster's transition to the DSC phase is not possible.

The comparison with MW GCs in \figref{fig:75} shows that almost all of them are located above the dashed curve, indicating that they cannot evolve into the DSC phase (as shown in \figref{fig:Rg-Size-2,3}). Therefore, assuming a canonical IMF and full initial retention of BHs, most MW GCs are nearly depleted of BHs, with only 0-1 per cent of their total mass comprising BHs. \figref{fig:75} shows that only about 13 per cent of MW GCs could currently contain 1-4 per cent of their mass in BHs. However, it indicates that approximately 85 per cent of their BHs have escaped since their formation. These GCs are located in the inner part of the MW (mean Galactocentric distance < 4$\kpc$) with an initial density $\log _{10}(\rho_\mathrm{h,i})<4.1$.

The magnitude of BH natal kick is a mystery that can be unlocked by examining the inferred present-day BH mass fraction within MW GCs. Recent studies have used different approaches to estimate this fraction, including best-fitting multimass models, evaluations of visible mass segregation, and assessment of the central surface brightness of several GCs. These studies indicate that typically, BH mass fractions ranging from 0-1 per cent of the GCs' total mass are needed to explain the observations, except for $\omega$ Cen \citep{Askar2018,Weatherford2020,Dickson2023}. According to \figref{fig:75}, even if we assume zero natal kicks for BHs, only 0-1 per cent of the present-day mass of the majority of MW GCs (87 per cent) consist of BHs. For the rest of the MW GCs, around 85 per cent of BHs have been ejected from the cluster up to the present-day. These results highlight that achieving the present-day BH mass fraction doesn't require BHs to receive a high natal kick. Instead, even with high initial retention of BHs, a substantial number of them are depleted through few-body encounters in the core of the GCs, shaping the present-day BH mass fraction.

According to \citet{Gieles2023}, the reduction of $\widetilde{\mathit{M}}_\mathrm{BH}(t)$ due to dynamical ejections depends only on the square root of the initial density relative to the tidal density before the cluster fills the tidal radius, and not on the initial relaxation timescale. Therefore, although increasing the initial modeled cluster mass primarily influences parameters such as the cluster relaxation time, it alone is unlikely to have a significant effect on $\widetilde{\mathit{M}}_\mathrm{BH}(t)$, as the $\widetilde{\mathit{M}}_\mathrm{BH}(t)$ of the cluster is mostly determined by cluster density. As a result, even though the data in \figref{fig:75} is specific to the modeled clusters with a mass of $3\times10^4 \Msun$, it could be extrapolated to more massive clusters similar to those of the MW GCs.

Note that Figures \ref{fig:Rg-Size-2,3} and \ref{fig:75} demonstrate the $\widetilde{\mathit{\tau}}_\mathrm{DSC}$ and fraction of BHs for clusters that initially retain all of their BHs. However, if we apply natal velocity kicks, such as canonical supernova kicks with mass fallback to the modeled clusters, leads to a re-scaling of the colors in these figures. Therefore, it is more accurate to state that Figures \ref{fig:Rg-Size-2,3} and \ref{fig:75} essentially depict $\widetilde{\mathit{\tau}}_\mathrm{DSC}$ and the fraction of BHs for clusters with $\widetilde{\mathit{M}}_\mathrm{BH}(0)=0.04$, irrespective of their IMF and the natal kick received by their BHs.

According to our simulations of clusters with a canonical IMF, when the cluster evolves into the DSC phase, $M_\mathrm{dyn}/M_\mathrm{photo}$ is approximately equal to 1.25, and its evolution is similar to $Q*$ and rises sharply in the DSC phase. Similarly, the dynamical mass-to-light ratio has the same evolution as $Q*$.

\subsection{Top-heavy IMF}\label{sec:top}

\begin{figure*}
    \centering
    \includegraphics[scale=1.0]{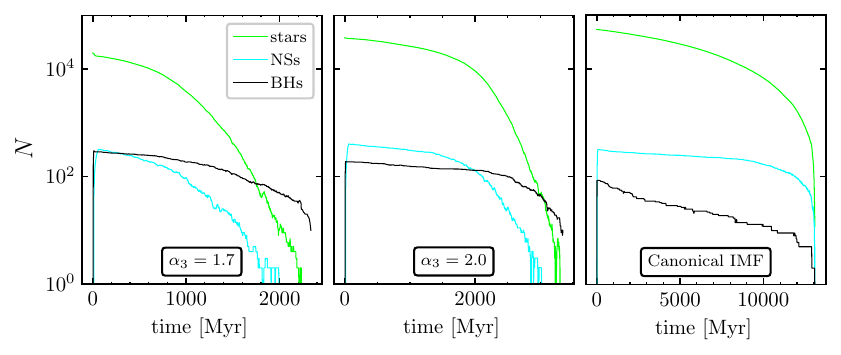} 
    \caption{The variation of the luminous star, neutron star, and black hole populations in clusters with different slopes of $\alpha_\mathrm{3}$ = 1.7, 2.0, and 2.3.}
    \label{fig:th}
\end{figure*}

\begin{figure}

    \includegraphics[scale=0.1818]{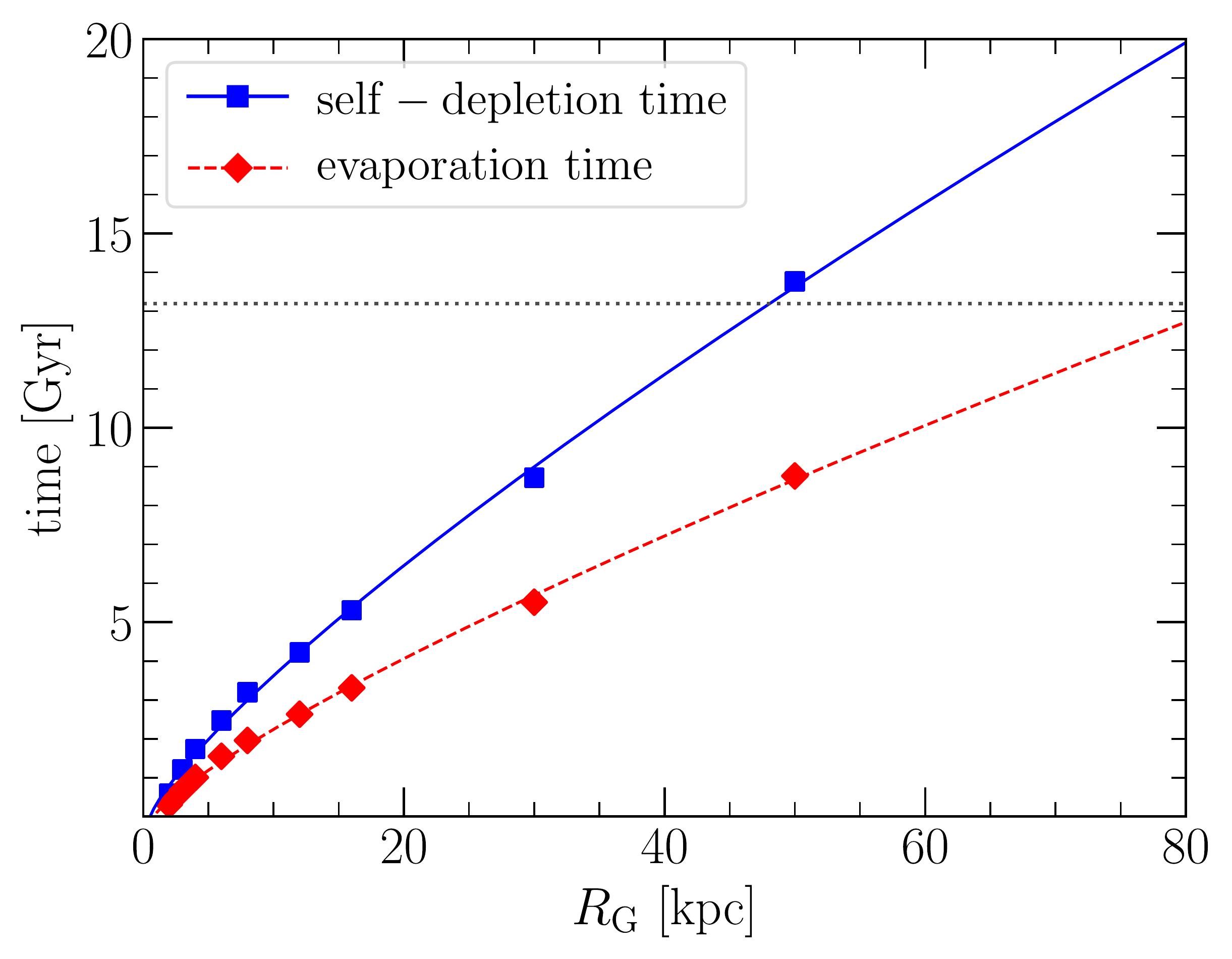}
    \includegraphics[scale=0.505]{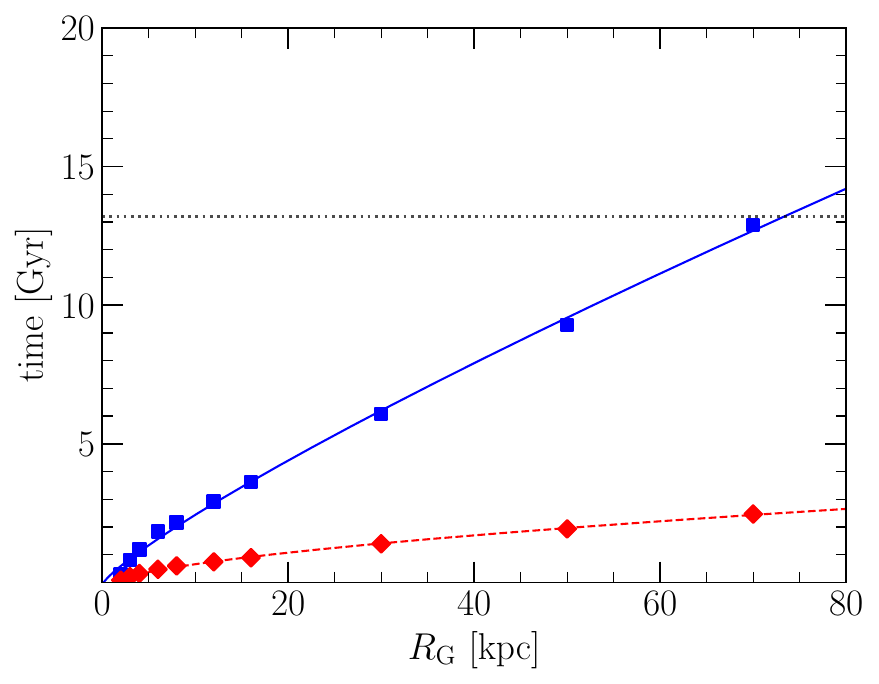}  
    \caption{Same as \figref{fig:rg}, but for clusters with $\alpha_\mathrm{3}$ = 2.0 (top panel) and $\alpha_\mathrm{3}$ = 1.7 (bottom panel).}
    \label{fig:dep-eva-top}
\end{figure}

\begin{figure}
\includegraphics[scale=0.43]{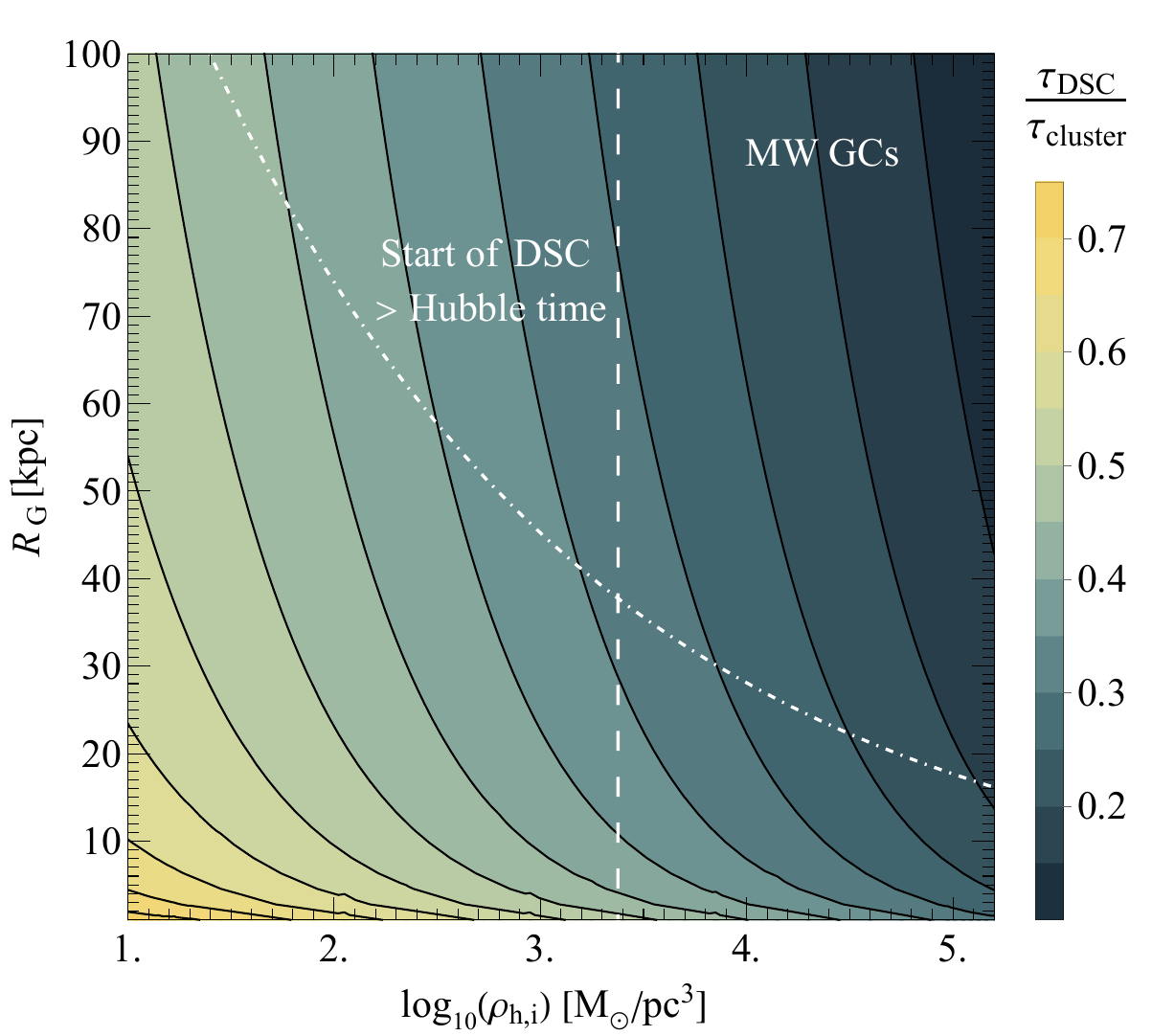}
\includegraphics[scale=0.43]{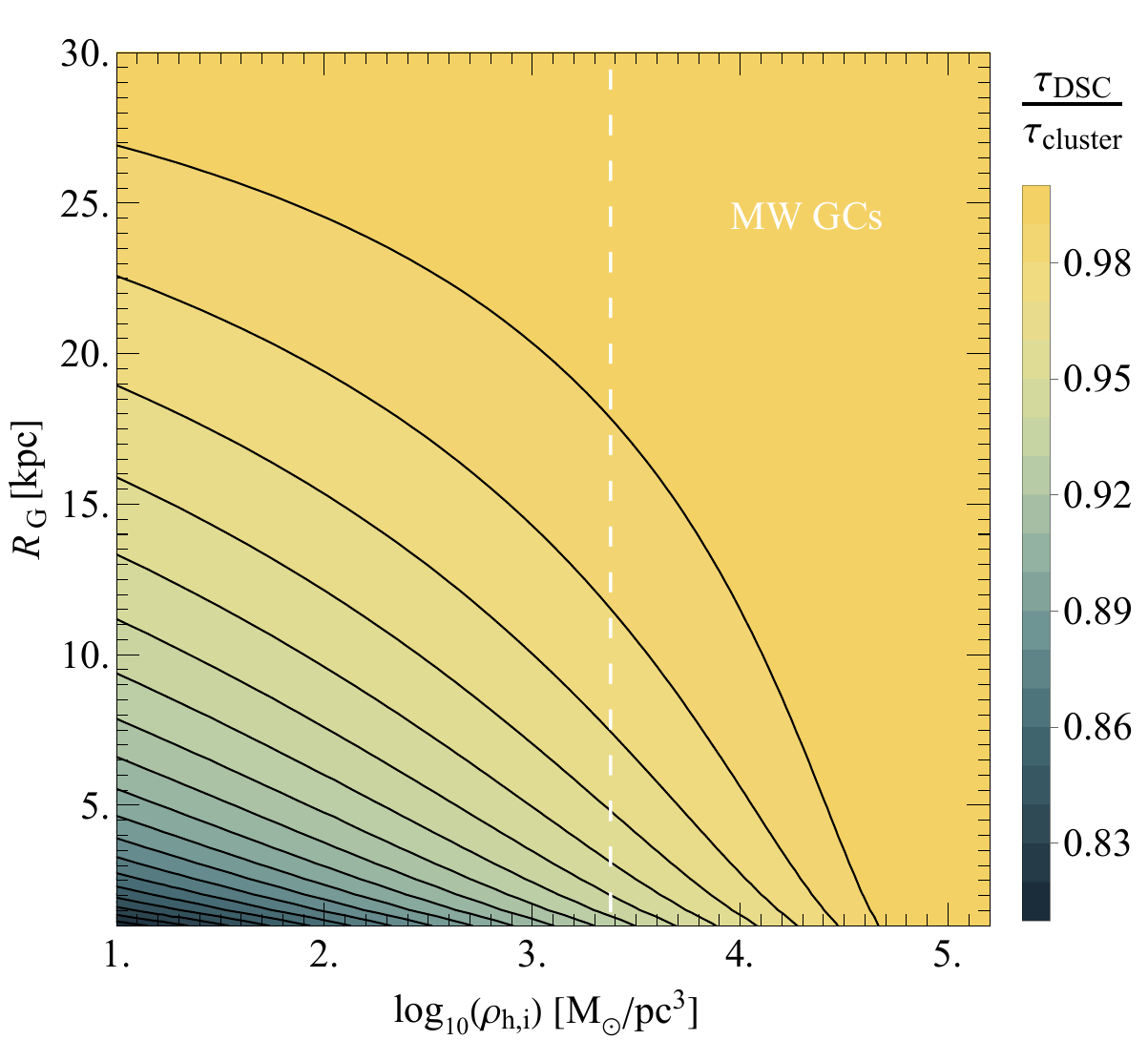} 
    \caption{Same as \figref{fig:Rg-Size-2,3}, but for clusters with $\alpha_\mathrm{3}$ = 2.0 (top panel) and $\alpha_\mathrm{3}$ = 1.7 (bottom panel).}
    \label{fig:Rg-Size-top}
\end{figure}

The number of BHs formed in a cluster is determined by the high-mass end of the stellar IMF. If the IMF is top-heavy, relatively more massive stars will be formed during star formation, resulting in a relatively greater number of BHs. This increase in BHs causes BH heating through mass segregation and consequently, a higher rate of few-body encounters in the BHSub, leading to a faster super-virial phase and increased evaporation rate of LSs. As a result, the dissolution timescale decreases with an increase in the slope of the mass function in high-mass stars, which is measured by the value of $\alpha_\mathrm{3}$ \citep{Chernoff1990,Chatterjee2017,Haghi2020, Weatherford2021}. To examine the effects of the IMF on the evolution of the DSC phase, we set up a series of models (sets B and C) with varying degrees of top-heaviness.

\figref{fig:th} depicts the time-evolution of the number of LSs (lime line), NSs (cyan line), and BHs (black line) in clusters with different values of $\alpha_\mathrm{3}$ that orbit circularly at a radius of $R_\mathrm{G}=8\kpc$. For a cluster with a canonical IMF (right panel), the mass fraction of BHs at birth is $\widetilde{\mathit{M}}_\mathrm{BH}(0)=0.04$. In this cluster, the energy produced by the BHSub is not significant enough to affect the evaporation rate of LSs, so the BHSub dissolves before the LSs have enough time to evaporate. As the BHSub self-depletes, the NSs form a segregated subsystem located at the centre of the cluster. As the cluster approaches complete dissolution, a short-lived DSC phase appears with $\widetilde{\mathit{\tau}}_\mathrm{DSC}=0.016$.

For a cluster with $\alpha_\mathrm{3}=2.0$ (middle panel), the mass fraction of BHs at birth is $\widetilde{\mathit{M}}_\mathrm{BH}(0)=0.09$. In this cluster, the BHSub produces enough energy to significantly increase the evaporation rate of LSs. Therefore, BHs remain in the cluster until the final stage of cluster evolution, while LSs and NSs have already evaporated. Consequently, the cluster spends approximately half of its lifetime in the DSC phase ($\widetilde{\mathit{\tau}}_\mathrm{DSC}=0.555$). Assuming a very top-heavy IMF with $\alpha_\mathrm{3}=1.7$ (left panel), the cluster becomes highly dominated by BHs ($\widetilde{\mathit{M}}_\mathrm{BH}(0)=0.15$), and the produced BHSub remains largely intact until dissolution, while other components have already escaped. In this cluster, which contains a significant number of BHs, the BHSub produces so much energy that LSs become super-virial already after approximately  100$\Myr$. This cluster is in the DSC phase for almost its entire life ($\widetilde{\mathit{\tau}}_\mathrm{DSC}=0.958$).

As illustrated in \figref{fig:dep-eva-top}, for the examined top-heavy models, $\alpha_\mathrm{3}=2.0$ (B1-B9) and $\alpha_\mathrm{3}=1.7$ (C5-C14), the self-depletion time is longer than the evaporation time at all Galactocentric distances. This means that all computed models can evolve to the DSC phase. For the $\alpha_\mathrm{3}=2.0$ models (\figref{fig:dep-eva-top} top panel), the difference between self-depletion time and evaporation time increases as $R_\mathrm{G}$ increases, while the parameters $(\tau_\mathrm{dep}-\tau_\mathrm{eva})/\tau_\mathrm{dep}$ and $\widetilde{\mathit{\tau}}_\mathrm{DSC}$ decrease slowly as $R_\mathrm{G}$ increases. It's worth noting that for clusters located above $\approx50\kpc$, the self-depletion time becomes longer than the Hubble time (shown as a dashed gray line in \figref{fig:dep-eva-top}). Clusters evolving at orbital radii larger than $80\kpc$ experience neither the LSs evaporation nor the DSC phase before the Hubble time.

The modeled clusters C5-C14 with $\alpha_\mathrm{3}=1.7$ generate intense energy from BH segregation, causing the LSs to become super-virial within approximately 100$\Myr$ and entering the DSC phase at the same time as the BHs segregate. The tidal field does not have a significant effect at this time. Even an isolated cluster (C16) enters the DSC phase around the same time. As the LSs become super-virial, they quickly evaporate from the cluster, and increasing $R_\mathrm{G}$ has little effect on reducing the evaporation rate. The bottom panel of \figref{fig:dep-eva-top} reveals that the evaporation time of LSs increases slowly with $R_\mathrm{G}$, but the self-depletion time increases rapidly. Once LSs escape, only the BHSub remains, with a longer dissolution time in weaker tidal fields (larger $R_\mathrm{G}$). Therefore both parameters, $(\tau_\mathrm{dep}-\tau_\mathrm{eva})/\tau_\mathrm{dep}$ and $\widetilde{\mathit{\tau}}_\mathrm{DSC}$, increase with increasing $R_\mathrm{G}$.

Analyzing the calculated models with various densities and Galactocentric distances, we discovered the most accurate linear relationship between $\widetilde{\mathit{\tau}}_\mathrm{DSC}$ and $\log _{10}(\rho_\mathrm{h,i})$ as well as $\log _{10}(R_\mathrm{G})$ (as seen in Equation \ref{eq:P2,3}). This relationship is depicted in \figref{fig:Rg-Size-top} for both $\alpha_\mathrm{3}=2.0$ (top panel) and $\alpha_\mathrm{3}=1.7$ (bottom panel). From \figref{fig:Rg-Size-top}, we can conclude that with a top-heavy IMF, many MW GCs can enter the DSC phase (for $\alpha_\mathrm{3}=2.0$, only clusters with $R_\mathrm{G}<35\kpc$) before the Hubble time.

A cluster's relaxation time decreases with a reduced $r_\mathrm{h,i}$, leading to a decrease in segregation time (Equation \ref{eq:mass_segregation}). Reduction in the segregation time of BHs can hasten the attainment of the super-virial state of LSs for clusters with $\alpha_\mathrm{3}=1.7$. For example, the DSC phase starts for C3, C11, and C19 ($R_\mathrm{G}=16\kpc$) at 53, 105, and 174$\Myr$, respectively. As the clusters reach the DSC phase more quickly, the $\widetilde{\mathit{\tau}}_\mathrm{DSC}$ increases. As a result, for clusters with $\alpha_\mathrm{3}=1.7$, the $\widetilde{\mathit{\tau}}_\mathrm{DSC}$ exhibits a positive correlation with $R_\mathrm{G}$ and $\rho_\mathrm{h,i}$, unlike as for the canonical IMF and top-heavy IMF with $\alpha_\mathrm{3}=2.0$. The bottom panel of \figref{fig:Rg-Size-top} highlights that the clusters with an initial density of MW GCs spend almost their entire lifetime in the DSC phase.

The results indicate that, when a cluster transitions into the DSC phase ($Q*=1$), the ratio of $M_\mathrm{dyn}/M_\mathrm{photo}$ is roughly 1.35 and 1.5 for $\alpha_\mathrm{3}=2.0$ and $\alpha_\mathrm{3}=1.7$, respectively. Clusters with a top-heavy IMF exhibit a softer evolution of $Q*$, $M_\mathrm{dyn}/M_\mathrm{photo}$, and $M_\mathrm{dyn}/L$ during the DSC phase, unlike the sharp increase observed for  the canonical IMF.

\subsection{The DSC transition boundary for different degrees of top-heaviness}\label{sec:DISCUSSION}

Based on the findings presented in \secref{sec:Canonical} and \secref{sec:top}, it is evident that the value of  $\widetilde{\mathit{\tau}}_\mathrm{DSC}$ relies on several factors, including $Z,\ \rho_\mathrm{h,i},\ R_\mathrm{G}\ \mathrm{and} \ \alpha_\mathrm{3}$ (i.e., $\widetilde{\mathit{M}}_\mathrm{BH}(0)$). The larger the value of $\widetilde{\mathit{\tau}}_\mathrm{DSC}$, the longer the cluster will remain in the BH-dominated state, leading to a more rapid increase in the $\widetilde{\mathit{M}}_\mathrm{BH}(t)$ ratio, which can eventually reach 1.0. Hence, if $\widetilde{\mathit{\tau}}_\mathrm{DSC}>0$, the $\widetilde{\mathit{M}}_\mathrm{BH}(t)$ ratio will continue to increase over time, and the transition boundary is defined by $\widetilde{\mathit{\tau}}_\mathrm{DSC}=0$, below which value  $\widetilde{\mathit{M}}_\mathrm{BH}(t)$ decreases over time, ultimately resulting in the complete depletion of BHs in the cluster before the cluster dissolves.

\begin{figure}

  \centering\includegraphics[scale=0.41]{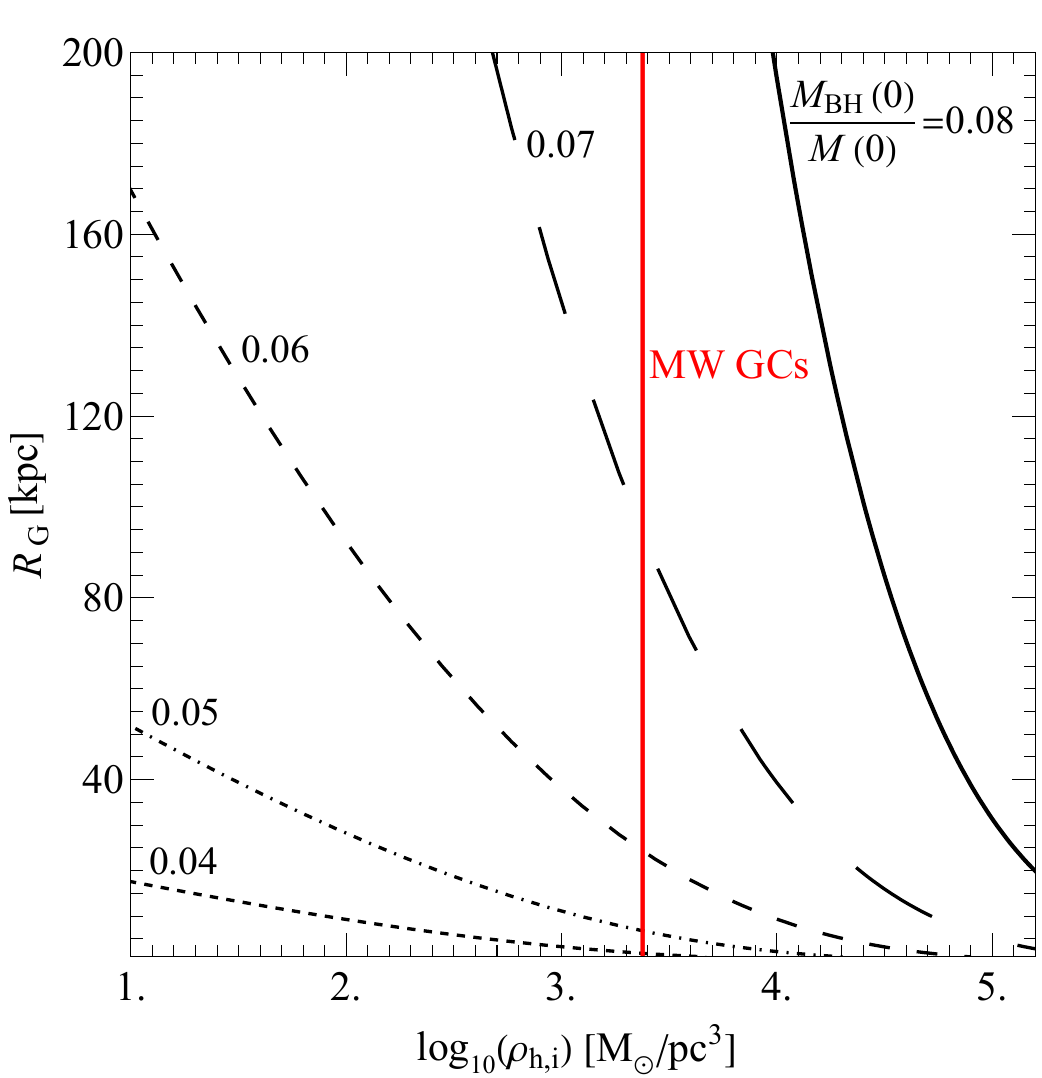}
  \caption{The contours where the $\widetilde{\mathit{\tau}}_\mathrm{DSC}$ is zero at $\widetilde{\mathit{M}}_\mathrm{BH}(0)=0.04, \ 0.05, \ 0.06,  \ 0.07, \ \mathrm{and} \ 0.08$ in the 2D space of $\rho_\mathrm{h,i}$ and $R_\mathrm{G}$. Star clusters located below the curves can evolve to the DSC phase.}
  \label{fig:Rg-Size-M}

\end{figure}

The transition boundary at which $\widetilde{\mathit{\tau}}_\mathrm{DSC}$ is zero for a given $\widetilde{\mathit{M}}_\mathrm{BH}(0)$ in $R_\mathrm{G}-\rho_\mathrm{h,i}$ space is plotted in \figref{fig:Rg-Size-M}. The contours show the necessary boundary conditions for a cluster to evolve into the DSC phase, based on a specific $\widetilde{\mathit{M}}_\mathrm{BH}(0)$ value. When the value of $\widetilde{\mathit{M}}_\mathrm{BH}(0)$ increases, the region in the $R_\mathrm{G}$ and $\log _{10}(\rho_\mathrm{h,i})$ space where clusters can transform into the DSC phase expands. \figref{fig:Rg-Size-M} shows that for MW GCs to evolve into the DSC phase, the retention fraction of BHs at birth must be greater than 0.05. Additionally, it can be concluded that the DSC phase will occur in all values of $\rho_\mathrm{h,i}$ and $R_\mathrm{G}$ when $\widetilde{\mathit{M}}_\mathrm{BH}(0)>0.08$. It is important to note that the limit of the Hubble time was not accounted for in \figref{fig:Rg-Size-M}. The determination of transition boundaries depends only on the initial mass fraction of BHs in the cluster, regardless of the magnitude of the natal kick received by them. It is expected that these transition boundaries will still be applicable for clusters that retain the mass fraction of BHs, $\widetilde{\mathit{M}}_\mathrm{BH}(0)$, even after experiencing the natal velocity kick of BHs. It is important to note that the region of space leading to a DSC is highly sensitive to $\widetilde{\mathit{M}}_\mathrm{BH}(0)$. Even canonical supernova kicks with mass fallback, let alone higher-speed kicks without fallback, reduce $\widetilde{\mathit{M}}_\mathrm{BH}(0)$ by about a factor of 2, which can alter the interpretation of the plot. This means that even realistic MW GCs with slightly top-heavy IMFs may not be capable of evolving to a DSC phase.

Recently, \citet{Zocchi2019} fitted dynamical models to $\omega$ Cen data and showed that models with 5 per cent of their mass in BHs can reproduce the data. Then \citet{Baumgardt2019} found by \Nbody simulation that a model containing 4.6 per cent of the cluster mass in a centrally concentrated cluster of stellar-mass BHs is a viable alternative to an Intermediate-Mass Black Hole (IMBH) model. According to our findings in Section \ref{sec:Canonical} ($\omega$ Cen is marked in \figref{fig:75}), the MW GCs with a canonical IMF cannot have 5 per cent of their mass in BHs at the present day and the BHSub is depleted from the cluster. This indicates that the IMF of $\omega$ Cen must have been top-heavy in order to have more than 5 per cent of its mass in the BHSub after 12$\Gyr$. Our findings are in agreement with \citet{MarksMichael2012}, who predicted a top-heavy IMF for $\omega$ Cen. Determining the mass fraction of BHs in GCs can thus be a method to estimate their IMF.

According to the analytical estimation of \citet{Breen2013}, the transition boundary for two-component systems is similar to our results. When the total mass fraction of BHs is small in tidally limited systems, meaning that $\widetilde{\mathit{M}}_\mathrm{BH}(0)<0.11$, the BHs deplete faster as their hosting clusters evolve, and $\widetilde{\mathit{M}}_\mathrm{BH}(t)$ decreases. Otherwise $\widetilde{\mathit{M}}_\mathrm{BH}(t)$ increases and finally a DSC forms. This transition boundary ($\widetilde{\mathit{M}}_\mathrm{BH}(0)=0.11$) is obtained based on the assumption that the mass loss of LSs and BHs both depend on the relaxation time and the total mass. Recent \Nbody simulations of star clusters with two components and no stellar evolution by \citet{Longwang2020} showed that in tidally filling models, $\widetilde{\mathit{M}}_\mathrm{BH}(t)$ increases rapidly in models with $\widetilde{\mathit{M}}_\mathrm{BH}(0)>0.07$, leading to the formation of BH-dominated dark clusters. However, the value of $0.07$ is not constant and depends on the ratio of the average mass of BHs to the average mass of other stars and the tidal radius divided by the half-mass radius. In our study, we performed 50 simulations that accounted for stellar evolution and varied initial conditions such as $\widetilde{\mathit{M}}_\mathrm{BH}(0)$, $\rho_\mathrm{h,i}$, and $R_\mathrm{G}$. Our results indicate that the transition boundary for a cluster to enter the DSC phase is approximately $\widetilde{\mathit{M}}_\mathrm{BH}(0)=0.08$.

\begin{figure}

 \centering\includegraphics[scale=0.46]{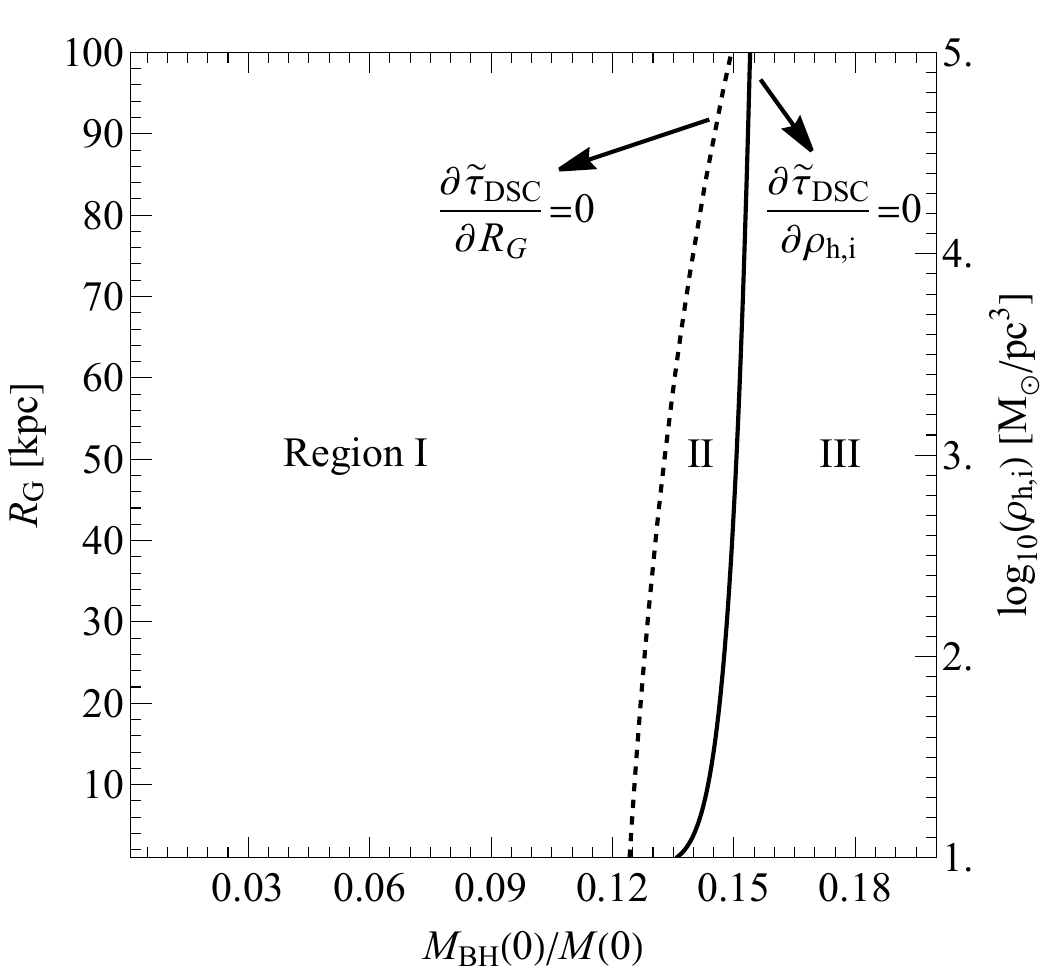}
 \caption{To determine the positive or negative correlation of $\widetilde{\mathit{\tau}}_\mathrm{DSC}$, the phase space is divided into three regions based on the variation of $R_\mathrm{G}$, $\rho_\mathrm{h,i}$, and $\widetilde{\mathit{M}}_\mathrm{BH}(0)= M_\mathrm{BH}(0)/M(0)$. The black curve indicates the points where $\partial \widetilde{\mathit{\tau}}_\mathrm{DSC}/\partial \rho_\mathrm{h,i}$ becomes zero for specific values of $R_\mathrm{G}$ and $\widetilde{\mathit{M}}_\mathrm{BH}(0)$. The dashed curve represents the points where $\partial \widetilde{\mathit{\tau}}_\mathrm{DSC}/\partial R_\mathrm{G}$ becomes zero for specific values of $\rho_\mathrm{h,i}$ and $\widetilde{\mathit{M}}_\mathrm{BH}(0)$. }
  \label{fig:Phase}

\end{figure}

According to the results obtained in the previous sections,  the dependency of $\widetilde{\mathit{\tau}}_\mathrm{DSC}$ on $R_\mathrm{G}$ and $\rho_\mathrm{h,i}$ as a function of $\widetilde{\mathit{M}}_\mathrm{BH}(0)$ can be obtained. The solid line in \figref{fig:Phase} separates the region in which $\partial \widetilde{\mathit{\tau}}_\mathrm{DSC}/\partial \rho_\mathrm{h,i}$ is positive (III) or negative (II, I). The dashed line in \figref{fig:Phase} separates the region in which $\partial \widetilde{\mathit{\tau}}_\mathrm{DSC}/\partial R_\mathrm{G}$  is positive (II and III) or negative (I). In between (II) is the region in which  $\partial \widetilde{\mathit{\tau}}_\mathrm{DSC}\left/\partial R_\mathrm{G}\right.>0$ and $\partial \widetilde{\mathit{\tau}}_\mathrm{DSC}/\partial \rho_\mathrm{h,i} <0$. As can be seen, for star clusters with  $\widetilde{\mathit{M}}_\mathrm{BH}(0)\leq0.12$,  $\widetilde{\mathit{\tau}}_\mathrm{DSC}$ decreases with increasing $R_\mathrm{G}$ and  $\rho_\mathrm{h,i}$, while it increases for star clusters with  $\widetilde{\mathit{M}}_\mathrm{BH}(0)\geq0.15$.   For clusters in this region, the energy generated from the segregation of BHs is so intense that the LSs become super-virial within nearly the first 100 Myr. Therefore, clusters with $\widetilde{\mathit{M}}_\mathrm{BH}(0)\geq0.15$ spend their entire lifetime in the DSC phase.

\section{CONCLUSION}\label{sec:conclusion}

Using numerical simulations of star clusters with a canonical IMF, \citet{banerjee2011}  predicted, for the first time, the formation of dark star clusters (DSC) as a result of BH segregation to the centre of the cluster and rapid removal of stars from the outer parts of a cluster caused by the strong tidal field of the host galaxy. As a generalization of that work, we explore the formation of DSCs in star clusters starting with a top-heavy IMF. In this work, we have carried out a series of direct \Nbody simulations of star clusters over a wide range of half-mass radii, metallicities, Galactocentric distances and IMF slope in the high-mass range to investigate the starting time and duration of the DSC phase (see \tabref{tab:initial_conditions} and \tabref{tab:initial_conditions_top_heavy}). In all simulations, we assumed zero natal kicks for NSs and BHs. With the complete retention fraction of the BHs, the cluster becomes Spitzer unstable, which leads to the formation of a black hole subsystem (BHSub) in the central part of the cluster.

The energy produced by the BHSub causes the LSs to enter a super-virial phase and speeds up their evaporation rate. Whether or not the cluster can reach the DSC phase depends on the time between the depletion of the BHSub and the evaporation of the LSs. We use the scaled DSC lifetime ($\widetilde{\mathit{\tau}}_\mathrm{DSC}$), which is defined as the duration of the cluster's DSC phase divided by its overall lifetime, to measure how long the BHSub dominates the evolution of the cluster. We examined the dependency of $\widetilde{\mathit{\tau}}_\mathrm{DSC}$ on $Z,\ \rho_\mathrm{h,i},\ R_\mathrm{G}\ \mathrm{and} \ \alpha_\mathrm{3}$ (or equivalently the mass fraction of the BHs at birth, $\widetilde{\mathit{M}}_\mathrm{BH}(0)$). The main outcomes of our study can be summarized as follows:

\begin{itemize}

\item The ratio of the dynamical mass (obtained from velocity dispersion) to the photometric mass ($M_\mathrm{dyn}/M_\mathrm{photo}$) and the dynamical mass-to-light ratio ($M_\mathrm{dyn}/L$) of the cluster are other parameters that can indicate the DSC phase. These observable parameters show a strong similarity to the $Q*$ parameter and undergo a similar change. We showed that at the time when the cluster evolves into the DSC phase (i.e., $Q*=1$), $M_\mathrm{dyn}/M_\mathrm{photo}$ is approximately equal to 1.25, 1.35, and 1.5 for $\alpha_\mathrm{3}=2.3,\ 2.0 \ \mathrm{and} \ 1.7$, respectively. Generally, a cluster is in the DSC phase if its dark remnant mass fraction is greater than 28 per cent assuming the WDs are luminous masses.

\item If the scaled DSC lifetime is equal to zero, then the ratio of the mass of the BHSub to the total mass of the cluster, $\widetilde{\mathit{M}}_\mathrm{BH}(t)$, decreases over time. This will result in the BHSub being depleted in the cluster. However, if $\widetilde{\mathit{\tau}}_\mathrm{DSC}>0$, then the $\widetilde{\mathit{M}}_\mathrm{BH}(t)$ ratio increases and eventually reaches 1.0.

\item The presence of a BHSub leads to the formation of a considerable amount of binaries containing BHs and BBHs. These are excellent sources of X-ray binaries and soft gravitational waves. The merging of BBHs, which results in detectable gravitational waves, mainly occurs during the initial stages of cluster evolution. However, if the cluster evolves into a DSC during the later stages of its evolution, it is unlikely to detect such a gravitational wave event during the DSC phase.

\item The lifetimes and scaled DSC lifetimes of the metal-rich ($Z=Z_\odot$) clusters are approximately two and a half times longer than those of the metal-poor ($Z=0.05Z_\odot$) clusters.

\item In order for the MW GCs to evolve into the DSC phase, the retention fraction of the BHs at birth must be greater than $0.05$. We have determined the minimum value of the initial BH mass fraction, $\widetilde{\mathit{M}}_\mathrm{BH}(0)$, that guarantees the cluster will reach the DSC phase. If this fraction is greater than $0.08$, the DSC phase will occur at every Galactocentric distance and with varying initial density.

\item If MW GCs have followed the canonical IMF and even if their BHs have not received any natal kicks, most of them are nearly depleted of BHs at present, with only 0-1 per cent of their total mass attributed to BHs. However, around 13 per cent of MW GCs could still have 1-4 per cent of their mass in BHs. Nevertheless, approximately 85 per cent of their BHs have escaped since their birth. These GCs are located near the bulge, within a mean Galactocentric distance of less than 4$\kpc$, and have a low initial density of less than $\log _{10}(\rho_\mathrm{h,i})<4.1$. Recent studies have shown that BH mass fractions ranging from 0-1 per cent of the total masses of MW GCs are typically required to explain the observations. Based on this, we can conclude that achieving the present-day BH mass fraction does not require BHs to receive a high natal kick. Even with high initial retention of BHs, a substantial number of them are depleted through few-body encounters in the core of the GCs, shaping the present-day BH mass fraction.

\item The identification of a BH mass fraction exceeding 1 per cent within MW GCs that orbit the Galactic center at mean Galactocentric distances larger than 4$\kpc$, or those characterized by high initial densities ($\log _{10}(\rho_\mathrm{h,i})>4.1$), can be interpreted as evidence for them having been born with a top-heavy IMF. The IMF of a GC can thus be constrained by determining the mass fraction of its BHs.

\item Several studies indicate that a model comprising over 5 per cent of the mass of $\omega \ \mathrm{Cen}$ in a cluster of BHs that are concentrated in the center could replicate the velocity dispersion profile of $\omega \ \mathrm{Cen}$ without requiring an IMBH at the center of the cluster. We showed that this suggests that the IMF of $\omega \ \mathrm{Cen}$ may have been top-heavy in order to have more than approximately 5 per cent of its mass in a BHSub after 12$\Gyr$.

\item  We showed that the scaled DSC lifetime, $\widetilde{\mathit{\tau}}_\mathrm{DSC}$, decreases with increasing the  Galactocentric distance and initial density of star clusters if $\widetilde{\mathit{M}}_\mathrm{BH}(0)\leq 0.12$,  while $\widetilde{\mathit{\tau}}_\mathrm{DSC}$ shows a positive correlation with both $R_\mathrm{G}$ and $\rho_\mathrm{h,i}$ if $\widetilde{\mathit{M}}_\mathrm{BH}(0)\geq 0.15$.

\end{itemize}

\section*{Acknowledgements}
AHZ and HH are grateful to the Helmholtz-Institut f\"ur Strahlen-und Kernphysik (HISKP), Universit\" at Bonn, for hospitality during their visit.  We thank the referee and the editor, for a constructive review.

\section*{Data availability}
The data underlying this article are available in the article.


\bibliographystyle{mnras}
\bibliography{references}

\bsp
\label{lastpage}
\end{document}